\documentclass[floats,aps,showpacs,reprint,pre]{revtex4-1}

\usepackage[dvips]{graphicx}
\usepackage{amsfonts}
\usepackage{amsmath,amssymb}
\usepackage{dsfont}
\usepackage{placeins}
\usepackage{afterpage}
\usepackage{flafter} 

\newcommand{\N}{\mathbb{N}}
\newcommand{\R}{\mathbb{R}}

\newcommand{\heff}{h_{\text{eff}}}
\newcommand{\hbareff}{\hbar_{\text{eff}}}

\newcommand{\A}{A}
\newcommand{\Areg}{A_{\text{reg}}}

\newcommand{\Achb}{\mathcal{A}_{\text{ch}}}

\newcommand{\Int}{\int\limits}
\newcommand{\ud}{\text{d}}

\newcommand{\monmat}{\mathcal{M}}

\newcommand{\bq}{\boldsymbol{q}}
\newcommand{\bp}{\boldsymbol{p}}

\newcommand{\psiregW}{\psi_{\text{reg},W}}

\newcommand{\ui}{\text{i}}
\newcommand{\ue}{\text{e}}
\newcommand{\psireg}{\psi_{\text{reg}}}
\newcommand{\psich}{\psi_{\text{ch}}}

\newcommand{\psicht}{\widetilde{\psi}_{\text{ch}}}
\newcommand{\Ureg}{U_{\text{reg}}}
\newcommand{\Uregt}{\widetilde{U}_{\text{reg}}}

\newcommand{\Preg}{P_{\text{reg}}}
\newcommand{\Pch}{P_{\text{ch}}}
\newcommand{\Pcht}{\widetilde{P}_{\text{ch}}}

\newcommand{\UT}{U_{\text{T}}}
\newcommand{\UVreg}{U_{\widetilde{V}}}
\newcommand{\UTreg}{U_{\widetilde{T}}}

\newcommand{\Hreg}{H_{\text{reg}}}
\newcommand{\Ereg}{E_{\text{reg}}}
\newcommand{\vchm}{v_{\text{ch},m}}
\newcommand{\Vchm}{V_{\text{ch},m}}

\newcommand{\Veffm}{V_{\text{eff}}^{m}}
\newcommand{\vchmn}{v_{\text{ch},mn}}
\newcommand{\Vchmn}{V_{\text{ch},mn}}
\newcommand{\mmax}{N_{\text{reg}}}
\newcommand{\opone}{\mathds{1}}
\newcommand{\phireg}{\varphi_{\text{reg}}}
\newcommand{\Omegareg}{\Omega_{\text{reg}}}

\newcommand{\no}{n_\Omega}

\newcommand{\real}{\text{Re}}
\newcommand{\imag}{\text{Im}}

\newcommand{\Qod}{Q_{\text{1d}}}
\newcommand{\Qdyn}{Q_{\text{dyn}}}
\newcommand{\Nch}{N_{\text{ch}}}
\newcommand{\mapho}{\mathcal{D}_{\text{ho}}}
\newcommand{\mapdef}{\mathcal{D}_{\text{d}}}
\newcommand{\mapwc}{\mathcal{D}_{\text{wc}}}
\newcommand{\mapd}{\mathcal{D}}
\newcommand{\mapsi}{\mathcal{D}_{\text{rs}}}
\newcommand{\xh}{\hat{x}}
\newcommand{\yh}{\hat{y}}
\newcommand{\km}{k_m}
\newcommand{\Zmn}{Z_{m}^{0}}

\begin{document}

\title{Direct regular-to-chaotic tunneling rates using the fictitious integrable system approach}

\author{Arnd B\"acker}
\author{Roland Ketzmerick}       
\author{Steffen L\"ock}
\affiliation{Institut f\"ur Theoretische Physik, Technische Universit\"at
             Dresden, 01062 Dresden, Germany}

\date{\today}

\begin{abstract}
We review the fictitious integrable system approach which predicts
dynamical tunneling rates from regular states to the chaotic region 
in systems with a mixed phase space. It is based on the introduction 
of a fictitious integrable system that resembles the regular dynamics within 
the regular island. We focus on the direct 
regular-to-chaotic tunneling process which dominates, if nonlinear resonances 
within the regular island are not relevant. For quantum maps, 
billiard systems, and optical microcavities we find excellent agreement 
with numerical rates for all regular states. 
\end{abstract}


\maketitle

\noindent


\section{Introduction}
\label{sec:intro}

Tunneling of a particle is one of the central manifestations of 
quantum mechanics. The prototypical example is the tunneling escape from a 
one-dimensional potential well through an energy barrier. While classically
the particle is confined for all times, quantum mechanically
the probability inside the well decays exponentially, $\exp(-\gamma t)$,
where $\gamma$ is the tunneling rate. It depends on the width and the height of 
the barrier and can be predicted, e.g., using WKB theory \cite{LanLif1991}.
Tunneling vanishes in the semiclassical limit, where typical classical actions
are large compared to Planck's constant.

Tunneling not only occurs for potential barriers but whenever the corresponding 
classical system consists of dynamically disconnected regions in phase space, 
which has been termed \textit{dynamical tunneling} \cite{DavHel1981}. 
It occurs in Hamiltonian systems which typically have a mixed phase
space. Here regions of regular motion, the so-called regular islands, and 
regions of chaotic motion, the so-called chaotic sea, coexist. 
While the classical motion is confined to any of these regions, 
quantum mechanically they are coupled by dynamical tunneling. 
In particular the fundamental process of \textit{regular-to-chaotic tunneling} 
describes the exponential decay of a wave packet 
initially localized in the regular island to the chaotic sea.
The same coupling also leads to tunneling from the chaotic sea to the 
regular island.

Dynamical tunneling also affects the structure of eigenstates of 
systems with a mixed phase space. According to the semiclassical eigenfunction
hypothesis \cite{Per1973,Ber1977,Vor1979} 
the eigenstates are concentrated 
either in the regular islands or in the chaotic sea. 
Away from the semiclassical limit  this classification still holds
approximately such that the corresponding 
eigenstates are called regular or chaotic. However, each eigenstate has 
contributions in the other regions of phase space, due to dynamical tunneling.

Dynamical tunneling in a mixed phase space has been studied theoretically 
\cite{HanOttAnt1984,Wil1986,BohTomUll1993,TomUll1994,Tom1998,Cre1998,ShuIke1995,
ShuIke1998,OniShuIkeTak2001,BohBooEgyMar1993,DorFri1995,FriDor1998,
BroSchUll2001,BroSchUll2002,EltSch2005,SchEltUll2005,WimSchEltBuc2006,
MouEltSch2006,Mou2007,DenMou2010,Kes2005,Kes2005b,PodNar2003,PodNar2005,
SheFisGuaReb2006,BarBet2007,BaeKetLoeSch2008,BaeKetLoeRobVidHoeKuhSto2008,
BaeKetLoeWieHen2009,LoeBaeKetSch2010,KesSch2011b}
and experimentally, e.g., in cold atom systems
\cite{SteOskRai2001,HenHafBroHecHelMcKMilPhiRolRubUpc2001,MouDel2003},
microwave billiards \cite{DemGraHeiHofRehRic2000,HofAltDemGraHarHeiRehRic2005,
BaeKetLoeRobVidHoeKuhSto2008}, and semiconductor nanostructures 
\cite{FroWilHayEavSheMIuHen2002}.
It is of current interest for, e.g., eigenstates affected by flooding of 
regular islands \cite{SchOttKetDit2001,BaeKetMon2005,BaeKetMon2007,Bit2010},
emission properties of optical microcavities 
\cite{WieHen2006,BaeKetLoeWieHen2009,ShiHarFukHenSasNar2010,
YanLeeMooLeeKimDaoLeeAn2010}, and spectral statistics in systems with a mixed 
phase space 
\cite{PodNar2007,VidStoRobKuhHoeGro2007,BatRob2010,BaeKetLoeMer2010}.

The concept of \textit{chaos-assisted tunneling} was introduced in 
Refs.~\cite{BohTomUll1993,TomUll1994,Tom1998}. It occurs, e.g., in
systems with two symmetry-related regular islands surrounded by chaotic
motion in phase space. There it was observed that the energy splittings 
$\Delta E$ between two symmetry related regular states are typically drastically 
enhanced due to the appearance of chaotic states, compared to an integrable 
situation. Chaos-assisted tunneling consists of two processes:
a regular-to-chaotic tunneling step from a regular torus of one island 
to the chaotic sea and a chaotic-to-regular tunneling step 
from the chaotic sea to the symmetry-related torus.
The energy splittings $\Delta E$ show strong fluctuations 
\cite{LinBal1990,BohTomUll1993,TomUll1994,Tom1998}
under variation of external parameters, as the distance between the 
energies of the regular doublet and the close-by chaotic states varies. 
This was also observed for optical microcavities 
\cite{PodNar2005} and microwave billiards 
\cite{DemGraHeiHofRehRic2000,HofAltDemGraHarHeiRehRic2005}.
In contrast to the energy splittings $\Delta E$ the regular-to-chaotic
tunneling rates $\gamma$ describe the average coupling to the chaotic sea,
$\gamma=\langle\Delta E\rangle/\hbar$, as shown in Sec.~\ref{sec:cat},
and therefore do not fluctuate. 

In the regime, $h\lesssim\Areg$, in which
the Planck constant $h$ is smaller but of the same order as the area
$\Areg$ of the regular island, quantum mechanics is not
affected by fine-scale structures in phase space, such as nonlinear
resonances or the hierarchical transition region. 
Hence, the \textit{direct regular-to-chaotic tunneling} process
dominates. In Ref.~\cite{HanOttAnt1984} a qualitative argument for the 
tunneling rates to behave exponentially as $\gamma \propto \exp(-B/h)$ 
was given. One may write $B=C\Areg$, where the non-universal constant  
$C$ has been calculated for various situations in the semiclassical limit: For a
weakly chaotic system $C = 2\pi$ \cite{SheFisGuaReb2006} was found and 
$C > 2\pi$ for rough nanowires 
\cite{FeiBaeKetRotHucBur2006,FeiBaeKetBurRot2009}. 
The approach in Ref.~\cite{PodNar2003} leads semiclassically to $C = 2 - \ln 4$ 
which was corrected to $C = 3 - \ln 4$ \cite{She2005}.
However, this prediction does not describe generically shaped regular
islands as it uses a transformation of the regular island to a harmonic 
oscillator.

In order to find a quantitative prediction of direct regular-to-chaotic
tunneling rates for generic islands we introduced the fictitious integrable 
system approach \cite{BaeKetLoeSch2008,Loe2010}. 
It relies on the decomposition of Hilbert space into a regular and a 
chaotic subspace by means of a fictitious integrable system 
\cite{BohTomUll1993,PodNar2003,SheFisGuaReb2006}. We require that its
dynamics resembles the regular motion in 
the originally mixed system as closely as possible and extends it beyond
the regular region. This leads to a tunneling
formula involving properties of this integrable system as
well as its difference to the mixed system under consideration.
It allows for the prediction of tunneling rates from
any quantized torus within the regular island. This approach was applied
to quantum maps \cite{BaeKetLoeSch2008}, 
billiard systems \cite{BaeKetLoeRobVidHoeKuhSto2008}, and optical
microcavities \cite{BaeKetLoeWieHen2009}.
 
In the semiclassical regime, $h\ll\Areg$, the fine-scale structures of the
classical phase space can be resolved by quantum mechanics. 
In particular, nonlinear
resonances cause an enhancement of the regular-to-chaotic tunneling rates, 
which has been termed \textit{resonance-assisted tunneling} 
\cite{BroSchUll2001,BroSchUll2002,EltSch2005,SchEltUll2005}. 
It leads to characteristic peak and plateau structures in the tunneling rates 
as observed for near integrable systems \cite{BroSchUll2001,BroSchUll2002}, 
mixed quantum maps \cite{EltSch2005,SchEltUll2005}, periodically driven systems 
\cite{MouEltSch2006,WimSchEltBuc2006}, quantum accelerator modes 
\cite{SheFisGuaReb2006}, and for multidimensional molecular 
systems \cite{Kes2005,Kes2005b}. Quantitatively, however, deviations of several 
orders of magnitude to numerical rates appear, especially in the experimentally
accessible regime where nonlinear resonances become relevant for tunneling. 
Recently, it was shown that a combination of 
the direct regular-to-chaotic tunneling mechanism and the resonance-assisted 
tunneling mechanism leads to a theory which quantitatively predicts 
tunneling rates from the quantum to the semiclassical regime 
\cite{LoeBaeKetSch2010}. For the application of this unified theory it is 
essential to determine the direct regular-to-chaotic tunneling rates.

In this paper we review the fictitious integrable system approach 
for the prediction of direct regular-to-chaotic tunneling rates. 
The approach is derived in Sec.~\ref{sec:deriv}.
Numerical methods for the determination 
of tunneling rates are presented in Sec.~\ref{sec:numerics}. 
It is then shown how the approach can be applied analytically, 
semiclassically, and numerically to quantum maps in Sec.~\ref{sec:maps:app} 
and two-dimensional billiard systems in Sec.~\ref{sec:billiards}.


\section{Dynamical tunneling and the fictitious integrable system approach}
\label{sec:deriv}

\begin{figure}[t]
  \begin{center}
    \includegraphics[width=8.5cm]{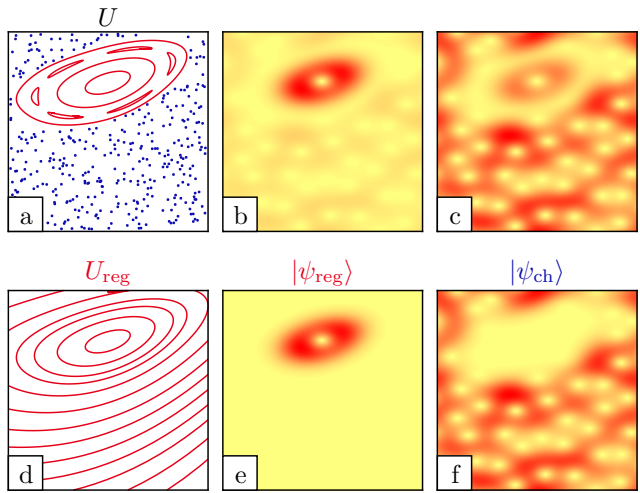}
    \caption{(Color online) (a) Illustration of the mixed classical phase 
          space corresponding to a quantum map $U$ 
          together with the Husimi representation of (b) a regular and (c)
          a chaotic eigenstate of $U$ which both have a small component in the
          other region. (d) Illustration of the classical 
          phase space of the fictitious integrable 
          system $\Ureg$. (e) Eigenstates $|\psireg\rangle$ 
          of $\Ureg$ are purely regular, while (f) the purely chaotic states 
          $|\psich\rangle$ extend in the chaotic region of phase space.
          }
    \label{fig:deriv:quantum_maps}
  \end{center}
\end{figure}

We consider systems with a mixed phase space, in particular two-dimensional 
quantum maps and billiards. Their phase space is divided into 
regions of regular dynamics and regions of chaotic 
dynamics. We focus on the fundamental situation of just one 
regular island embedded in
the chaotic sea, Fig.~\ref{fig:deriv:quantum_maps}(a). 
At the center of the island one has an elliptic fixed point, which is 
surrounded by invariant regular tori.
For such systems the semiclassical eigenfunction hypothesis 
\cite{Per1973,Ber1977,Vor1979} implies that in the semiclassical
limit the eigenstates can be classified
as either regular or chaotic, according to the phase-space
region on which they concentrate.
In order to understand the behavior of eigenstates 
away from the semiclassical limit one has to compare the size of
phase-space structures with Planck's constant $h$. 
We discuss this exemplarily for quantum 
maps \cite{BerBalTabVor1979} which are described by a unitary time-evolution 
operator $U$ on a Hilbert space of finite dimension $N$. Here we introduce the
semiclassical parameter $\heff=h/\A=1/N$ as the ratio of Planck's constant $h$ 
to the area $\A$ of phase space. The
eigenstates $|\psi_n\rangle$ and quasi-energies
$\varphi_n$ of $U$ are determined by
\begin{eqnarray}
\label{eq:deriv:eigen_eq}
 U|\psi_n\rangle = \ue^{\ui\varphi_n}|\psi_n\rangle.
\end{eqnarray}
The so-called regular states are predominantly concentrated on tori 
within the regular island and fulfill the Bohr-Sommerfeld type 
quantization condition
\begin{eqnarray}
\label{eq:deriv:EBK}
 \oint p\,\ud q = \heff\left(m+\frac{1}{2}\right),\quad m=0,\dots,\mmax-1.
\end{eqnarray}
For a given value of $\heff$ there exist $\mmax$ of such 
regular states, where $\mmax=\lfloor \Areg/\heff+1/2 \rfloor$ and $\Areg$ 
from now on is the dimensionless area of the regular island. 
The chaotic states mainly extend over the chaotic sea. 
Note, that for systems with a large density of chaotic states the 
regular states may disappear and chaotic states flood the regular island 
\cite{BaeKetMon2005,BaeKetMon2007}.

An important consequence of a finite $\heff$ in systems with a
mixed phase space is dynamical tunneling. It couples the regular island and the 
chaotic sea, which are classically separated. 
Hence, the regular and chaotic eigenfunctions of $U$ always 
have a small component in the other region of phase space, respectively,
see Fig.~\ref{fig:deriv:quantum_maps}(b,c). 

The coupling of the regular and the chaotic phase-space regions can be 
quantified by tunneling rates $\gamma_m$ which describe the decay from the 
$m$th regular torus to the chaotic sea. To define these tunneling rates one 
can consider a wave packet started on this $m$th quantized torus in the regular 
island which is coupled to a continuum of chaotic states, as in the case of an 
infinite chaotic sea. Its decay
$\ue^{-\gamma_m t}$ is characterized by a tunneling rate $\gamma_m$.
For systems with a finite phase space this exponential decay occurs at most up 
to the Heisenberg time $\tau_H=\heff/\Delta_{\text{ch}}$, where 
$\Delta_{\text{ch}}$ is the mean level spacing of the chaotic states. 
Alternatively, the tunneling rates can be obtained from lifetimes of 
resonances in a corresponding open system, e.g., by adding an absorbing region 
somewhere in the chaotic component, see Sec.~\ref{sec:numerics:open}.

In the regime, $\heff \lesssim \Areg$, where $\heff$ is smaller but 
comparable to the area $\Areg$ of the regular island, the rates 
$\gamma_m$ are dominated by the direct regular-to-chaotic tunneling mechanism, 
while contributions from resonance-assisted tunneling are negligible.
We concentrate on systems where additional phase-space structures 
within the chaotic sea are not relevant for tunneling.
In the following we derive a prediction for the direct regular-to-chaotic 
tunneling rates using the fictitious integrable system approach
\cite{BaeKetLoeSch2008}.

\subsection{Derivation}
\label{sec:deriv:2}

In order to find a prediction for the direct regular-to-chaotic tunneling rates
we decompose the Hilbert space of the quantum map $U$ into two parts which
correspond to the regular and chaotic regions.
While classically such a decomposition is unique, 
quantum mechanically this is not the case due to the uncertainty 
principle. We find a decomposition by introducing a fictitious integrable 
system $\Ureg$. Related ideas were presented in 
Refs.~\cite{BohTomUll1993,PodNar2003,SheFisGuaReb2006}. The fictitious 
integrable system has to be chosen such that its dynamics resembles 
the classical motion corresponding to $U$ within the regular island as closely 
as possible and continues this regular dynamics beyond the regular island of 
$U$, see Fig.~\ref{fig:deriv:quantum_maps}(d). 
The eigenstates $|\psireg^m\rangle$ of $\Ureg$, 
$\Ureg|\psireg^m\rangle = \ue^{\ui \phireg^m}|\psireg^m\rangle$, 
are purely regular in 
the sense that they are localized on the $m$th quantized torus 
of the regular region and continue to decay beyond this regular region, 
see Fig.~\ref{fig:deriv:quantum_maps}(e). This is the decisive property of
$|\psireg^m\rangle$ which have no chaotic admixture, 
in contrast to the predominantly regular eigenstates of $U$, 
see Fig.~\ref{fig:deriv:quantum_maps}(b). 
The explicit construction of $\Ureg$ is 
discussed in Sec.~\ref{sec:regsyst}.

With the eigenstates $|\psireg^m\rangle$ of $\Ureg$ we define a projection 
operator
\begin{equation}
\label{eq:deriv:P}
 \Preg := \sum_{m=0}^{\mmax-1} |\psireg^m\rangle\langle\psireg^m|,
\end{equation}
using the first $\mmax$ eigenstates of $\Ureg$ which approximately projects onto 
the regular island corresponding to $U$. 
The orthogonal projector 
\begin{equation}
\label{eq:deriv:Pch}
 \Pch := \opone-\Preg 
\end{equation}
approximately projects onto the chaotic phase-space region. These projectors, 
$\Preg$ and $\Pch$, define our decomposition of the Hilbert space into a regular
and a chaotic subspace.
 
Introducing a basis $|\psich\rangle$ in the chaotic subspace we can write
$\Pch = \sum_{\text{ch}} |\psich\rangle\langle\psich|$. Here we sum over all 
$N_{\text{ch}}=N-\mmax$ states $|\psich\rangle$, which we call purely chaotic 
states, see Fig.~\ref{fig:deriv:quantum_maps}(f) for an illustration.
The coupling matrix element $\vchm$ between a purely regular state 
$|\psireg^m\rangle$ and any purely chaotic state $|\psich\rangle$ is
\begin{equation}
\label{eq:deriv:coupmatelorth}
 \vchm = \langle\psich|U|\psireg^m\rangle.
\end{equation}
From this the tunneling rate is obtained using a dimensionless version of 
Fermi's golden rule, see Appendix~\ref{sec:appendix:fgr},
\begin{eqnarray}
\label{eq:deriv:FGR}
 \gamma_m = \sum_{\text{ch}}|\vchm|^2,
\end{eqnarray}
where the sum is over all chaotic basis states $|\psich\rangle$
and thus averages the modulus squared of the fluctuating matrix elements $\vchm$. 
Here we apply Fermi's golden rule in the case of a discrete spectrum, 
which is possible if one considers the decay $\ue^{-\gamma_m t}$
up to the Heisenberg time $\tau_H=\heff/\Delta_{\text{ch}}$ only.

Inserting Eq.~\eqref{eq:deriv:coupmatelorth} into Eq.~\eqref{eq:deriv:FGR} 
we obtain
\begin{eqnarray}
\label{eq:deriv:final_result}
 \gamma_m = \Vert\Pch U|\psireg^m\rangle\Vert^2 = 
            \Vert (\opone-\Preg)U|\psireg^m\rangle\Vert^2
\end{eqnarray}
as the basis of all our following investigations. 
It allows for the prediction of tunneling rates 
from a regular state localized on the $m$th quantized torus to the chaotic sea.
Equation~\eqref{eq:deriv:final_result} agrees with the intuition that the 
tunneling rates are determined by the amount of probability that is transferred 
to the chaotic region after one application of the time evolution operator $U$
on $|\psireg^m\rangle$.
We want to emphasize that Eq.~\eqref{eq:deriv:final_result} essentially relies
on the chosen decomposition of Hilbert space determined by the fictitious 
integrable system $\Ureg$. A similar expression for the tunneling rates was 
obtained from a phenomenological Hamiltonian in Ref.~\cite{SheFisGuaReb2006}.
Note, that the tunneling rate for the inverse process of tunneling from the 
chaotic sea to the $m$th regular torus is also given by 
Eq.~\eqref{eq:deriv:final_result} but with an additional prefactor 
of $1/\Nch$ due to the different density of final states \cite{Bit2010}.

\subsubsection{Approximation for very good $\Ureg$}
\label{sec:deriv:approx1}

In cases where one finds a fictitious integrable system $\Ureg$ which
resembles the dynamics within the regular island of $U$ with very high accuracy, 
Eq.~\eqref{eq:deriv:final_result} can be approximated as
\begin{eqnarray}
\label{eq:deriv:final_result_app}
 \gamma_m & \approx & \Vert (U-\Ureg)|\psireg^m\rangle \Vert^2,
\end{eqnarray}
using $\Preg U|\psireg^m\rangle \approx \Preg\Ureg|\psireg^m\rangle=
\Ureg|\psireg^m\rangle$.
Instead of the projector $\Pch$ in Eq.~\eqref{eq:deriv:final_result}
the difference $U-\Ureg$ enters in Eq.~\eqref{eq:deriv:final_result_app}.
This allows for a semiclassical evaluation, which is presented in 
Sec.~\ref{sec:maps:sc}.

\subsubsection{Approximation for non-orthogonal chaotic states}
\label{sec:deriv:approx2}

If one constructs chaotic states $|\psicht\rangle$ from random wave
models \cite{Ber1977}, they will not be orthogonal to the purely regular 
states $|\psireg^m\rangle$.
In this case we construct orthonormalized states
\begin{eqnarray}
|\psich\rangle & := & c\,(\opone-\Preg)|\psicht\rangle
\end{eqnarray}
with normalization $c$.
We find for the coupling matrix elements, Eq.~\eqref{eq:deriv:coupmatelorth},
\begin{eqnarray}
\label{eq:deriv:coupmatel3}   
 \vchm & = & c \langle\psicht|U-\Preg U |\psireg^m\rangle \\
\label{eq:deriv:coupmatel5}   
 & \approx & \langle\psicht|U-\Ureg|\psireg^m\rangle,
\end{eqnarray}
where we use the approximations $c\approx 1$ and again
$\Preg U |\psireg^m\rangle \approx \Ureg |\psireg^m\rangle$. 
Equation~\eqref{eq:deriv:coupmatel5} can now be inserted into 
Eq.~\eqref{eq:deriv:FGR}, leading to
\begin{eqnarray}
\label{eq:deriv:FGR_Ureg}
 \gamma_m & \approx & \Vert\Pcht(U-\Ureg)|\psireg^m\rangle\Vert^2
\end{eqnarray}
with $\Pcht=\sum_{\text{ch}}|\psicht\rangle\langle\psicht|$.

\subsubsection{Application to billiards}
\label{sec:deriv:billiard}

Two-dimensional billiard systems, which we consider in Sec.~\ref{sec:billiards}, 
are given by the motion of a free particle
of mass $M$ in a domain $\Omega$ with elastic reflections at its boundary 
$\partial\Omega$. Quantum mechanically they are described by a Hamilton 
operator $H$. The fictitious integrable system approach can also be applied 
to billiards: We use a
fictitious integrable system $\Hreg$ and its eigenstates $\psireg^{mn}(\bq)$,
characterized by the two quantum numbers $m$ and $n$. 
We start from a random wave model \cite{Ber1977} for the chaotic states 
$\psicht(\bq)$ which 
are not orthogonal to the purely regular states. Using the approximation 
for non-orthogonal chaotic states we obtain in analogy to 
Eq.~\eqref{eq:deriv:coupmatel5}
\begin{equation}
\label{eq:deriv:coupmatelorth_bil}
 \Vchmn = \Int_{\Omega} \ud^2q\, \psicht(\bq) (H-\Hreg) \psireg^{mn}(\bq)
\end{equation}
for the coupling matrix element between a purely regular state with quantum 
numbers $(m,n)$ and different chaotic states $\psicht(\bq)$.
The tunneling rate $\Gamma_{mn}$ is obtained using Fermi's golden rule,
Eq.~\eqref{eq:app:fgr:1}, 
\begin{eqnarray}
\label{eq:deriv:FGR_bil}
 \Gamma_{mn} = \frac{2\pi}{\hbar} \langle|\Vchmn|^2\rangle\,\rho_{\text{ch}} 
               \approx \frac{\Achb\hbar}{4M}\langle|\Vchmn|^2\rangle
\end{eqnarray}
where we average over the modulus squared of coupling matrix elements $\Vchmn$ 
between one particular purely regular state and different chaotic states 
of similar energy. The chaotic density of states $\rho_{\text{ch}}$ 
is approximated by the leading Weyl term 
$\rho_{\text{ch}}\approx\Achb\hbar^2/(8\pi M)$ in which $\Achb$ 
denotes the area of the billiard times the chaotic fraction of phase space.
This expression for $\rho_{\text{ch}} $ follows, e.g., from counting the number 
of Planck cells $h^2$ in the chaotic part of phase space \cite{BohTomUll1993}.

\subsection{Fictitious integrable system}
\label{sec:regsyst}

The most difficult step in the application of  
Eqs.~\eqref{eq:deriv:final_result} and \eqref{eq:deriv:FGR_bil} to a given 
system is the determination of the fictitious integrable system $\Hreg$. 
Its dynamics should resemble the classical motion 
of the considered mixed system within the regular island as closely as possible.
As a result the contour lines of the corresponding integrable Hamiltonian 
$\Hreg$, Fig.~\ref{fig:deriv:quantum_maps}(d), approximate the KAM-curves of the 
classically mixed system, Fig.~\ref{fig:deriv:quantum_maps}(a), in phase space.
This resemblance is not possible with arbitrary precision as the integrable 
approximation for example does not contain nonlinear resonance chains and small 
embedded chaotic regions. Moreover, it cannot account for the hierarchical 
regular-to-chaotic transition region at the border of the regular island.
Similar problems appear for the analytic continuation of a regular torus into 
complex space due to the existence of natural boundaries 
\cite{Wil1986,Cre1998,ShuIke1995,ShuIke1998,OniShuIkeTak2001,BroSchUll2001,
BroSchUll2002,EltSch2005}.
However, for not too small $\heff$, where 
these small structures are not yet resolved quantum mechanically, 
an integrable approximation with finite accuracy turns out to be sufficient 
for a prediction of the tunneling rates.

In addition the integrable dynamics of $\Hreg$ should extrapolate 
smoothly beyond the regular island of $H$. 
This is essential for the quantum eigenstates of $\Hreg$ to have correctly 
decaying tunneling tails. According to Eq.~\eqref{eq:deriv:final_result} they 
are relevant for the determination of the tunneling rates. 
While typically tunneling from the regular island occurs to 
regions within the chaotic sea close to the border of the regular island, there 
exist other cases, where it occurs to regions deeper inside the chaotic sea, as
studied in Ref.~\cite{SheFisGuaReb2006}. Here $\Hreg$ has to be constructed such 
that its eigenstates have the correct tunneling tails up to this region,
see Sec.~\ref{sec:maps:amphib:wc}.

For quantum maps we determine the fictitious integrable system in the following 
way: We employ classical methods, see below, to obtain a one-dimensional 
time-independent Hamiltonian $\Hreg(q,p)$
which is integrable by definition and resembles the classically regular motion
of the mixed system. After its quantization we obtain the regular quantum map 
$\Ureg=\ue^{-\ui\Hreg/\hbareff}$ with corresponding
eigenfunctions $|\psireg^m\rangle$. For the numerical evaluation of  
Eq.~\eqref{eq:deriv:final_result} we use 
$\Pch = \opone - P_{\text{reg}}=\opone-\sum|\psireg^m\rangle\langle\psireg^m|$, 
where the sum extends over $m=0,\,1,\,\dots,\,\mmax-1$.

Now we discuss two examples for the explicit construction of $\Hreg$. 
Note, that also other methods, e.g., based on the normal-form analysis 
\cite{Gus1966,BazGioSerTodTur1993} 
or on the Campbell-Baker-Hausdorff formula \cite{Sch1988} 
can be employed in order to find $\Hreg$. For the example systems considered in 
this paper, however, they show less good agreement.

\subsubsection{Lie-transformation method}
\label{sec:regsyst:lie}

\begin{figure}[tb]
  \begin{center}
    \includegraphics[width=8.5cm]{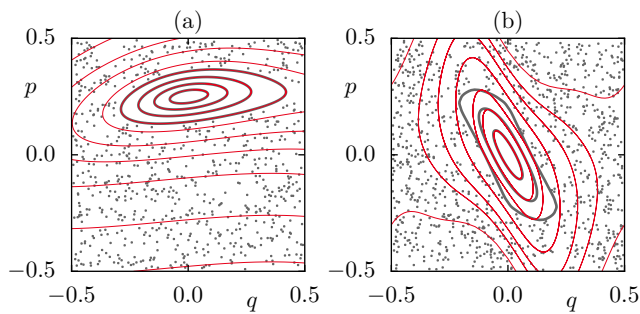}
    \caption{(Color online) Application of the Lie-transformation method. 
            (a) Orbits (thick gray lines and dots) 
            of the map $\mapdef$ (see Sec.~\ref{sec:maps:amphib:def}) 
            and of the corresponding 
            integrable system (thin red lines) of order $K=15$. Here $\Hreg$ 
            accurately resembles the regular dynamics of $U$. 
            (b) Orbits of the standard map (see Sec.~\ref{sec:maps:stmap})
            for $\kappa=2.9$ (thick gray lines and dots) and of the 
            corresponding integrable system (thin red lines) of order $K=7$.
            Here $\Hreg$ does not accurately resemble the regular dynamics 
            of $H$.}
    \label{fig:maps:comp_map_Hreg_lie}
  \end{center}
\end{figure}

One approach for the determination of the fictitious integrable system
for quantum maps is the Lie-transformation method 
\cite{LicLie1983}. It determines a classical Hamilton function
\begin{equation}
 \label{eq:deriv:hamiltonian_lie}
  \Hreg^K(q,p) = \sum_{l=0}^{K} \tau^l h_{l}(q,p)
\end{equation}
as a power series in the period of the driving $\tau$, see 
Fig.~\ref{fig:maps:comp_map_Hreg_lie}(a) and Ref.~\cite{BroSchUll2002}  
for examples.
Typically, the order of the expansion $K$ can be increased up to $20$ within 
reasonable numerical effort. The Lie-transformation method provides a
regular approximation $\Hreg$ which interpolates the dynamics inside the regular 
region and gives a smooth continuation into the chaotic sea. At some order
$K$ the series typically diverges due to the nonlinear resonances
inside the regular island. 
For strongly driven systems, such as the standard map at $\kappa>2.5$,
the Lie-transformation method is not able to reproduce the regular dynamics 
of $U$, see Fig.~\ref{fig:maps:comp_map_Hreg_lie}(b).

\subsubsection{Method using the frequency map analysis}
\label{sec:regsyst:fma}

\begin{figure}[tb]
  \begin{center}
    \includegraphics[width=8.5cm]{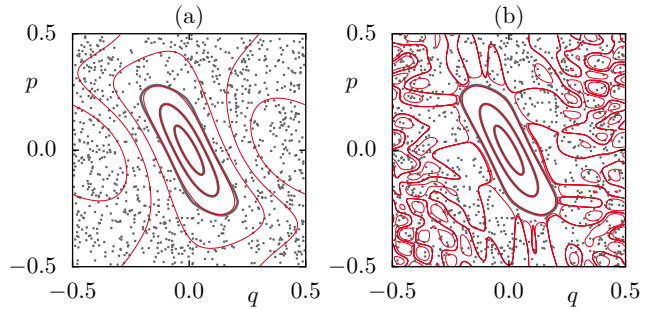}
    \caption{(Color online) 
             Application of the method using the frequency map analysis. 
             We show orbits of the standard map (see Sec.~\ref{sec:maps:stmap})
             for $\kappa=2.9$ (thick gray lines and dots)
             and the corresponding integrable system 
             (thin red lines) of order (a) $K=2$
             and (b) $K=10$. While in (b) $\Hreg$ resembles the regular island of
             $U$ with higher accuracy than in (a), the extrapolation of $\Hreg$ 
             beyond the island strongly oscillates.}
    \label{fig:maps:comp_map_Hreg_fma}
  \end{center}
\end{figure}

An alternative method is applicable even to strongly driven 
one-dimensional systems. 
In order to determine $\Hreg(q,p)$ we associate to each torus within the 
regular region of $U$ an energy. This information for individual tori is then 
extrapolated to the entire phase space. 
To this end we consider a straight line $(q(u),p(u))$, parametrized by $u$, 
from the center $u=0$ of the regular island to its border with the chaotic sea. 
Each torus of the map crosses this line at some value $u$ and using the frequency 
map analysis \cite{LasFroCel1992} we compute the enclosed area $A(u)$ and the 
rotation number $\nu(u)$.  
Using a polynomial interpolation of these functions we calculate an energy
\begin{eqnarray}
 E(u) = \Int_{0}^{u} \ud u' \, \nu(u') \frac{\ud A(u')}{ \ud u'}
\end{eqnarray}
for each torus in the regular region of phase space \cite{Loe2010}. 
This formula follows from 
Hamilton's equations of motion and $A(u)=\oint p(q,u)\,\ud q$.
Finally, we find the fictitious integrable system $\Hreg$ by a two-dimensional 
extrapolation of the energies to the whole phase space with
\begin{equation}
 \label{eq:deriv:hamiltonian_fma}
  \Hreg^K(q,p) = \sum_{k,l=-K}^{K} h_{k,l} \ue^{2\pi\ui kq} \ue^{2\pi\ui lp}
\end{equation}
using periodic basis functions up to the maximal order $K$.
An example is shown in Fig.~\ref{fig:maps:comp_map_Hreg_fma}(a).
Note, that the resulting $\Hreg$ shows a reasonable behavior beyond the 
regular island of $H$ only for small values of $K$. For too large orders $K$ 
one observes that $\Hreg$ oscillates
in this region, see Fig.~\ref{fig:maps:comp_map_Hreg_fma}(b).
This would lead to purely regular states $|\psireg^m\rangle$ with incorrect
tunneling tails beyond the regular island of $U$, resulting in wrong predictions
of tunneling rates with Eq.~\eqref{eq:deriv:final_result}. 

\subsubsection{Quality of the prediction}
\label{sec:converg}

An important question is whether the direct tunneling rates obtained
using Eq.~\eqref{eq:deriv:final_result}
depend on the actual choice of $\Hreg$ and how these results 
converge in dependence of the order $K$ of its perturbation series.
There are two main problems: First the classical expansion of the 
Lie transformation is asymptotically divergent \cite{LicLie1983}, 
which means that from some order $K$ on the series fails 
in reproducing the dynamics of the mixed system inside the regular 
region, see Fig.~\ref{fig:maps:comp_map_Hreg_lie}(b). 
Second, for the quantization of $\Hreg$ its behavior in the 
vicinity of the last surviving KAM torus must be smoothly continued beyond the 
regular island of $U$. Large fluctuations of $\Hreg$ in this region, as appear 
for the method based on the frequency map analysis for large $K$, make 
the use of Eq.~\eqref{eq:deriv:final_result} impossible,
see Fig.~\ref{fig:maps:comp_map_Hreg_fma}(b). 

Ideally one would like to use classical measures, which describe the 
deviations of the 
regular system $\Hreg$ from the originally mixed one, to predict the error of 
Eq.~\eqref{eq:deriv:final_result} for the tunneling rates. 
However, these classical measures can only account 
for the deviations within the regular region but not for the quality of the 
continuation of $\Hreg$ beyond the regular island of $U$. 
It remains an open question how to obtain a direct connection between the 
error on the classical side and the one for the tunneling rates. 

\begin{figure}[tb]
  \begin{center}
    \includegraphics[width=8.5cm]{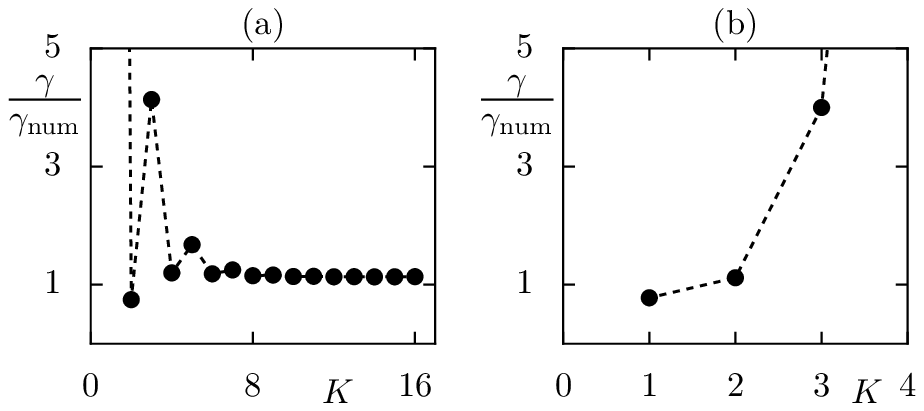}
    \caption{(Color online) Tunneling rates $\gamma$ determined 
             with Eq.~\eqref{eq:deriv:final_result}, 
             normalized by the numerical value $\gamma_{\text{num}}$ 
             for $m = 0$, $\heff = 1/32$ vs order 
             $K$ of $\Hreg$. In (a) we choose the system $\mapdef$ 
             (see Sec.~\ref{sec:maps:amphib:def}) and use the 
             Lie transformation for the
             determination of $\Hreg$ while in (b) the standard map 
             (see Sec.~\ref{sec:maps:stmap}) at $\kappa=2.9$
             is considered for which we determine $\Hreg$ using the method
             based on the frequency map analysis. 
          }
    \label{fig:maps:convergence}
  \end{center}
\end{figure}

Nevertheless, the convergence of the integrable approximation can be studied by 
considering the tunneling rates determined with
Eq.~\eqref{eq:deriv:final_result} under variation of 
the perturbation order $K$. 
For the example map  $\mapdef$ (introduced in Sec.~\ref{sec:maps:amphib:def})
we find convergence up to the maximal considered 
order, Fig.~\ref{fig:maps:convergence}(a), and later use $K=10$ for the comparison 
of Eq.~\eqref{eq:deriv:final_result} and numerical rates. For the 
standard map at $\kappa=2.9$ the rates diverge rather quickly, 
Fig.~\ref{fig:maps:convergence}(b), and we use $K=2$.

In general, different classical methods are applicable to determine a
fictitious integrable system $\Ureg$ which leads to an accurate prediction of
tunneling rates with Eq.~\eqref{eq:deriv:final_result}. Hence,
the determination of $\Ureg$ is not unique. The quality of an  
integrable system can be estimated a posteriori by comparison of the predicted 
tunneling rates with numerical rates.

\subsubsection{Application to billiards}

There are only few integrable two-dimensional billiard systems 
such as the circular, the rectangular, and the elliptical billiard. 
We use such integrable systems 
for various applications, as discussed in Sec.~\ref{sec:billiards}
for the mushroom and the annular billiard as well as for wires in a magnetic 
field and the annular microcavity.
A general procedure to obtain $\Hreg$ for arbitrary 
billiards is still under development.

\subsection{Semiclassical evaluation}
\label{sec:maps:sc}

In this section we semiclassically evaluate the direct regular-to-chaotic
tunneling rates for systems where the fictitious integrable system is of the 
form $\Hreg(q,p)=p^2/2+W(q)$.
We consider one-dimensional kicked systems
\begin{eqnarray}
\label{eq:maps:kicked_hamiltonian}
 H(q,p,t) = T(p) + V(q) \sum_{n}\tau \delta(t - n\tau)
\end{eqnarray}
which are the simplest Hamiltonian systems showing a mixed phase-space 
structure. They are described by the kinetic energy $T(p)$ and the potential 
$V(q)$ which is applied once per kick period $\tau=1$.
The classical dynamics of a kicked system is given by its stroboscopic mapping,
e.g., evaluated just after each kick
\begin{eqnarray}
\label{eq:maps:kicked_map}
 \begin{array}{l}
  q_{n+1}=q_{n}+T'(p_{n}),\\ 
  p_{n+1}=p_{n}-V'(q_{n+1}).
 \end{array}
\end{eqnarray}
It maps the phase-space coordinates after the $n$th kick to those
after the $(n+1)$th kick.
The corresponding quantum map over one kicking period is given by $U=U_V U_T$ 
with
\begin{eqnarray}  
\label{eq:maps:quantum_map}
 U_V & = & \ue^{-\ui V(q)/\hbareff}, \\
 U_T & = & \ue^{-\ui T(p)/\hbareff}.
\end{eqnarray}
We consider a compact phase space with periodic 
boundary conditions for $q \in [-1/2, \, 1/2]$ and $p \in [-1/2, \, 1/2]$. 
  
For an analytical evaluation of Eq.~\eqref{eq:deriv:final_result}, which 
predicts the direct regular-to-chaotic 
tunneling rates, we approximate the fictitious integrable system $\Ureg$ by
a kicked system, $\Uregt = \UVreg U_T$ or $\Uregt = U_V\UTreg$ with
\begin{eqnarray}  
\label{eq:maps:quantum_map_reg}
 \UVreg & = & \ue^{-\ui \widetilde{V}(q)/\hbareff}, \\
 \UTreg & = & \ue^{-\ui \widetilde{T}(p)/\hbareff}.
\end{eqnarray}
Here the functions $\widetilde{V}(q)$ and $\widetilde{T}(p)$ are a low
order Taylor expansion of $V(q)$ and $T(p)$, respectively, around the center 
of the regular island.
Note, that the classical dynamics corresponding to $\Uregt$ is typically not 
completely regular. Still the following evaluation is applicable if
$\Uregt$ has the properties:
(i) Within the regular island it has an almost identical classical dynamics 
as $U$, including nonlinear resonances and small embedded chaotic regions.
(ii) It shows predominantly regular dynamics for a sufficiently wide region 
beyond the border of the regular island of $U$. 

We now give a semiclassical evaluation of Eq.~\eqref{eq:deriv:final_result}
assuming that both properties (i) and (ii) are fulfilled. 
We consider the first case $\Uregt = \UVreg U_T$. 
As the dynamics of $U$ and $\Uregt$ are almost 
identical within the regular island of $U$, the approximate result, 
Eq.~\eqref{eq:deriv:final_result_app}, can be applied with $\Ureg$
replaced by $\Uregt$, giving
\begin{eqnarray}
\label{eq:deriv:final_result_app_2}
 \gamma_m & \approx & \Vert (U - \UVreg U_T)|\psireg^m\rangle\Vert^2 \\
          & = & \Vert (U U_{T}^{\dagger}\UVreg^{\dagger}-\opone) 
                \UVreg U_T |\psireg^m\rangle\Vert^2.
\end{eqnarray}
We now use that $|\psireg^m\rangle$, which is an eigenstate of the exact $\Ureg$
and $\Hreg$, is an approximate eigenstate of $\UVreg U_T$, leading to 
$\UVreg U_T|\psireg^m\rangle \approx 
\ue^{\ui \varphi_{\text{reg}}^{m}}|\psireg^m\rangle$.
We obtain 
\begin{eqnarray}
\label{eq:maps:sc:tunneling_rate_1b}
\gamma_m & \approx & \Vert (U_V \UVreg^{\dagger}-\opone) 
                     | \psireg^m \rangle\Vert^2. 
\end{eqnarray}
In position representation this reads
\begin{eqnarray}
  \label{eq:maps:sc:tunneling_rate_2}
    \gamma_m & \approx &
      2 \sum_{k=0}^{N-1} \left\vert \psireg^m (q_k)  \right
      \vert^{2} \left[ 1-\cos\left( \frac{\Delta V(q_k)}{\hbareff}
      \right)  \right],
\end{eqnarray}
where $\Delta V(q):=V(q)-\widetilde{V}(q)$ and $q_k = k/N-1/2$.
In the semiclassical limit the sum in Eq.~\eqref{eq:maps:sc:tunneling_rate_2} 
can be replaced by an integral over the position space
\begin{eqnarray}
  \label{eq:maps:sc:tunneling_rate_2b}
    \gamma_m & \approx & 
      2 \Int_{-1/2}^{1/2} \ud q \left\vert \psireg^m (q)  \right
      \vert^{2} \left[ 1-\cos\left( \frac{\Delta V(q)}{\hbareff}
      \right)  \right].
\end{eqnarray}
Here, for the normalization $\int \ud q \, \vert \psireg^m (q)  
\vert^{2}=1$ holds, while previously $\sum_k \vert 
\psireg^m (q_k) \vert^{2}=1$ was fulfilled.
Note, that for the second case, where $\Ureg\approx U_V\UTreg$ is used in 
Eq.~\eqref{eq:deriv:final_result_app}, a similar result can be obtained 
in momentum representation,
\begin{eqnarray}
  \label{eq:maps:sc:tunneling_rate_2c}
    \gamma_m & \approx & 
      2 \Int_{-1/2}^{1/2} \ud p \left\vert \psireg^m (p)  \right
      \vert^{2} \left[ 1-\cos\left( \frac{\Delta T(p)}{\hbareff}
      \right)  \right],
\end{eqnarray}
with $\Delta T(p):=T(p)-\widetilde{T}(p)$.

We now use a WKB expression for the regular states $|\psireg^m\rangle$. 
For simplicity we restrict to the case 
\begin{equation}
 \label{eq:maps:sc:wkb_wf_1d_hamiltonian}
 \Hreg(q,p) = \frac{p^2}{2} + W(q)
\end{equation}
leading to
\begin{eqnarray}
 \label{eq:maps:sc:wkb_wf_1d_ort}
  \psireg^m(q) & \approx & \sqrt{\frac{\omega}{2\pi|p(q)|}} 
                  \exp\left(-\frac{1}{\hbareff}\Int_{q_{m}^{r}}^{q} 
                  |p(q')|\text{d}q' \right),\;\;
\end{eqnarray}
which is valid for $q>q_{m}^{r}$.
Here $q_{m}^{r}$ is the right classical turning point of the $m$th quantizing 
torus, $\omega$ is the oscillation frequency, and $p(q)=\sqrt{2(\Ereg^m-W(q))}$.
The eigenstates $\psireg^m(q)$ decay exponentially beyond the classical turning 
point $q_{m}^{r}$. The difference of the potential energies $\Delta V(q)$ 
approximately vanishes within the regular region and increases beyond its border
to the chaotic sea. Hence, the most 
important contribution in Eq.~\eqref{eq:maps:sc:tunneling_rate_2b} arises 
near the left or the right border, $q_{b}^{l}$ or $q_{b}^{r}$, of the regular island. 
For $q>q_{b}^{r}$ we rewrite the regular states 
\begin{eqnarray}
\psireg^m(q) & \approx & \psireg^m\left(q_{b}^{r}\right) 
                  \exp\left(\frac{-1}{\hbareff}\Int_{q_{b}^{r}}^{q} 
                   |p(q')|\,\ud q'\right)
                  \sqrt{\tfrac{p(q_{b}^{r})}{p(q)}}\;\;\;\;\;\\
\label{eq:maps:sc:WKB_eval_3}
           & \approx & \psireg^m\left(q_{b}^{r}\right) 
             \exp\left(-\frac{1}{\hbareff}
             (q-q_{b}^{r})|p(q_{b}^{r})|\right),
\end{eqnarray}
where in the last step we use $p(q)\approx p(q_{b}^{r})$ in the vicinity of the
border.

In order to evaluate Eq.~\eqref{eq:maps:sc:tunneling_rate_2b} we split the integration 
interval into two parts, such that $\gamma_m = \gamma_{m}^{l}+\gamma_{m}^{r}$,
corresponding to the contributions from the left and the right.
For simplicity we now approximate $\Delta V(q)$ by a piecewise linear function,
\begin{eqnarray}
 \Delta V(q) \approx \left\{
 \begin{array}{ll}
   0 & ,\; q_{m}^{r} \leq q \leq q_{b}^{r} \\
   c_b (q-q_{b}^{r}) & ,\; q > q_{b}^{r}
 \end{array}
 \right.,
\end{eqnarray}
with a constant $c_b$. With this we find
\begin{eqnarray}
\label{eq:maps:sc:tunnelrate_randformel_intnorm_a}
\gamma_{m}^{r} & \approx & 2\hbareff \vert\psireg^m(q_{b}^{r})\vert^2 
             \Int_{0}^{x_{\text{max}}} \ue^{-2x|p(q_{b}^{r})|} 
             [1-\cos(c_b x)]\,\ud x\quad\;\;\\
\label{eq:maps:sc:tunnelrate_randformel_intnorm}
       & \approx & \frac{I\heff}{\pi} \vert\psireg^m(q_{b}^{r})\vert^2, 
\end{eqnarray}
where $x = (q-q_{b}^{r})/\hbareff$,
$x_{\text{max}} = (1/2-q_{b}^{r})/\hbareff$, and 
\begin{eqnarray}
\label{eq:maps:sc:tunnelrate_randformel_integral}
I = \Int_{0}^{x_{\text{max}}} \ue^{-2x|p(q_{b}^{r})|} [1-\cos(c_b x)]\,\ud x.
\end{eqnarray} 
In the semiclassical limit $x_{\text{max}}\to\infty$ and for fixed quantum 
number $m$ the integral $I$
becomes an $\heff$-independent constant. The tunneling rate $\gamma_{m}^{r}$ is 
proportional to the square of the modulus of the regular wave 
function at the right border $q_{b}^{r}$ of the regular island. 
With Eq.~\eqref{eq:maps:sc:wkb_wf_1d_ort} we obtain
\begin{eqnarray}
\label{eq:maps:sc:tunnelrate_rand_wkb_einges}
 \gamma_{m}^{r} & \approx & \frac{I\omega \heff}{2\pi^2\vert p(q_{b}^{r})\vert} 
              \exp\left(-\frac{2}{\hbareff} \Int_{q_{m}^{r}}^{q_{b}^{r}} 
              \left\vert p(q') \right\vert \ud q'\right).
\end{eqnarray}
A similar equation holds for $\gamma_{m}^{l}$.
Note, that the same exponent is obtained when considering the one-dimensional 
tunneling problem through an energy barrier in between the right turning point 
$q_{m}^{r}$ and the right border of the island $q_{b}^{r}$. 

As an example for the explicit evaluation of 
Eq.~\eqref{eq:maps:sc:tunnelrate_rand_wkb_einges} 
we consider the harmonic oscillator $\Hreg(q,p) = p^2/2+\omega^2q^2/2$,
where $\omega$ denotes the oscillation frequency and gives the ratio of the 
two half axes of the elliptic invariant tori.
Its classical turning points $q_{m}^{r,l} = \pm \sqrt{2E_m}/\omega$, the 
eigenenergies $E_m = \hbareff\omega(m+1/2)$, and the momentum 
$p(q) = \sqrt{2E_m-q^2\omega^2}$ are explicitly given.
Using these expressions in Eq.~\eqref{eq:maps:sc:tunnelrate_rand_wkb_einges} 
and $\gamma_m = 2\gamma_{m}^{r}$ we obtain
\begin{equation}
\label{eq:maps:sc:wkb_formula}
   \gamma_{m} =  
              c\,\frac{\heff}{\beta_m} \;
              \text{exp}\Bigg(-\displaystyle\frac{2\Areg}{\heff}\left[\beta_m - 
              \alpha_m \ln \left(
              \frac{1+\beta_m}{\sqrt{\alpha_m}}\right)\right]\Bigg)
\end{equation}
as the semiclassical prediction for the tunneling rate of the $m$th regular 
state, where $\Areg$ is the area of the regular island, 
$\alpha_m=(m+1/2) (\Areg/\heff)^{-1}$, and $\beta_m = \sqrt{1-\alpha_m}$. 
The exponent in
Eq.~\eqref{eq:maps:sc:wkb_formula} was also derived in
Ref.~\cite{DenMou2010} using complex-time
path integrals. The prefactor 
\begin{equation}
\label{eq:maps:sc:wkb_formula_prefactor}
 c = \frac{I}{\pi^2}\sqrt{\frac{\pi\omega}{\Areg}} 
\end{equation}
can be estimated semiclassically by solving the integral,
Eq.~\eqref{eq:maps:sc:tunnelrate_randformel_integral}, for $x_{\text{max}}\to\infty$.
For a fixed classical torus of energy $E$ one obtains
\begin{eqnarray}
\label{eq:maps:sc:wkb_formula_prefactor_int}
  I & \approx & \frac{1}{2|p(q_{b}^{r})|}-
                \frac{2|p(q_{b}^{r})|}{4|p(q_{b}^{r})|^2+c_b^2}.
\end{eqnarray}
With this prefactor the prediction 
Eq.~\eqref{eq:maps:sc:wkb_formula} gives excellent agreement 
with numerically determined rates over $10$ orders of magnitude in $\gamma$, 
see Fig.~\ref{fig:maps:rates_amph_ellipt}(c).
For a fixed quantum number $m$ in the semiclassical limit the energy $E_m$ 
approaches zero such that one can approximate 
$|p(q_{b}^{r})|\approx\omega q_{b}^{r}$
in Eq.~\eqref{eq:maps:sc:wkb_formula_prefactor_int} which does not depend on 
$\heff$.

Let us make the following remarks concerning Eq.~\eqref{eq:maps:sc:wkb_formula}:
The only information about this non-generic island with constant
rotation number is $\Areg/\heff$ as in Ref.~\cite{PodNar2003}. In contrast to 
Eq.~\eqref{eq:deriv:final_result} it does not require 
further quantum information such as the quantum map $U$.
While the term in square brackets semiclassically approaches one, 
it is relevant for large $\heff$.
In contrast to Eq.~\eqref{eq:maps:sc:tunneling_rate_2b}, where the chaotic
properties are contained in the difference $\Delta V(q)$, 
they now appear in the prefactor $c$ via the linear 
approximation of this difference.

In the semiclassical limit the tunneling rates 
predicted by Eq.~\eqref{eq:maps:sc:wkb_formula}
decrease exponentially. For $\heff\to 0$ one has $\alpha_m \to 0$ and
$\beta_m \to 1$, such that $\gamma \sim \ue^{-2\Areg/\heff}$. This reproduces the 
qualitative prediction obtained in Ref.~\cite{HanOttAnt1984}. The
non-universal constant in the exponent is $2$ which is comparable to the prefactor 
$3-\ln 4\approx 1.61$ derived in Refs.~\cite{PodNar2003,She2005}. 
We find that our result gives more accurate agreement 
to numerical rates, as will be shown in Sec.~\ref{sec:maps:app}. 
Still, a semiclassical evaluation
of Eq.~\eqref{eq:deriv:final_result} for a fictitious integrable system of a 
more general form than Eq.~\eqref{eq:maps:sc:wkb_wf_1d_hamiltonian}
has to be developed.

\subsection{Relation to chaos-assisted tunneling}
\label{sec:cat}

In Ref.~\cite{TomUll1994} Tomsovic and Ullmo studied dynamical tunneling in 
systems with two symmetry-related regular islands surrounded by a chaotic
region in phase space. They considered the quasi-energy splittings 
$\Delta \varphi_m$ between the symmetric and antisymmetric regular states 
on the $m$th quantizing tori of both islands.
These tunneling splittings are drastically 
enhanced by the presence of chaos, i.e.\ chaotic states assist the tunneling 
process compared to the case of a system with 
integrable dynamics between the regular islands.
The two-step process, which couples the regular 
torus from one island to the chaotic sea and from the chaotic sea to the 
symmetry-related torus of the other island, 
dominates the direct coupling of the two regular tori. 

The tunneling splittings $\Delta\varphi_m$ show fluctuations over several 
orders of magnitude under variation of external parameters 
\cite{LinBal1990,BohTomUll1993,TomUll1994}. These fluctuations originate from 
the varying distance of the regular doublet to the chaotic states and 
their varying coupling. According to a random matrix model,
the splittings follow a Cauchy distribution 
\cite{LeyUll1996} with geometric mean \cite{SchEltUll2005}
\begin{equation}
\label{eq:app:split:1}
 \langle\Delta \varphi_{m}\rangle = 
               \left(\frac{\sqrt{\Nch}\Veffm\tau}{\hbar}\right)^2,
\end{equation}
where $\Veffm$ describes the effective coupling of the $m$th regular state
to the chaotic sea and $\tau$ is the period of the driving.
Note, that the factor $\sqrt{\Nch}$ arises due to the different 
convention in Ref.~\cite[Eq.~(1.27)]{SchEltUll2005}, where the regular state
is coupled to one chaotic state only. 

We now show that this average tunneling splitting 
$\langle\Delta \varphi_{m}\rangle$ in systems with symmetry-related regular 
regions is identical to the tunneling rate $\gamma_m$
from one regular region to the chaotic sea:
We start from Eq.~\eqref{eq:app:fgr:2}, 
$\gamma_m=\Nch\langle|\vchm|^2\rangle$,
and use the relation of the dimensionless coupling matrix elements $\vchm$, 
defined in Eq.~\eqref{eq:deriv:coupmatelorth}, to $\Veffm$
\begin{equation}
\label{eq:app:split:3}
  \langle|\vchm|^2\rangle=\left(\frac{\Veffm\tau}{\hbar}\right)^2.
\end{equation}
Together with Eq.~\eqref{eq:app:split:1} this leads to
\begin{equation}
\label{eq:app:split:2}
 \gamma_m = \langle\Delta \varphi_{m}\rangle.
\end{equation}
This result was previously employed in Fig.~3 in 
Ref.~\cite{LoeBaeKetSch2010}, where numerical splittings $\Delta \varphi_{m}$ 
are used and the prediction is for tunneling rates $\gamma_m$.

\begin{figure}[tb]
  \begin{center}
    \includegraphics[width=8.5cm]{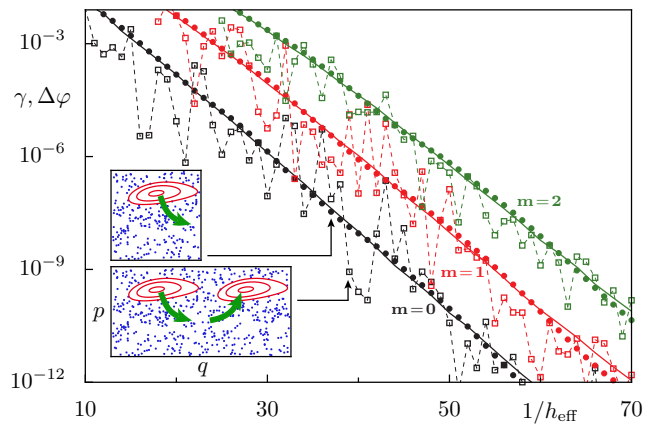}
    \caption{(Color online) Dynamical tunneling rates $\gamma$ (dots),
          quasi-energy splittings $\Delta\varphi$ (squares), 
          and the prediction of Eq.~\eqref{eq:deriv:final_result} (lines)
          vs $1/\heff$ for the regular states $m\leq 2$.
          We use the system $\mapdef$ (see Sec.~\ref{sec:maps:amphib:def}).
          The insets show the phase space with two symmetry-related 
          regular regions (used to determine $\Delta\varphi$) and the phase 
          space with one regular region (used to determine $\gamma$).
         }
    \label{fig:deriv:cat}
  \end{center}
\end{figure}

Fig.~\ref{fig:deriv:cat} illustrates the strong fluctuations of the splittings 
$\Delta\varphi_m$ (squares) in contrast to the smooth behavior of the 
tunneling rates $\gamma_m$ (dots, lines). As predicted by 
Eq.~\eqref{eq:app:split:2} one can see in the figure that the splittings 
fluctuate around the tunneling rates as a function of $1/\heff$. 

This demonstrates that for a quantitative verification of a theory on 
regular-to-chaotic tunneling the tunneling rates allow for a more precise
comparison than tunneling splittings.


\section{Numerical determination of tunneling rates}
\label{sec:numerics}

\begin{figure}[tb]
  \begin{center}
    \includegraphics[width=8.5cm]{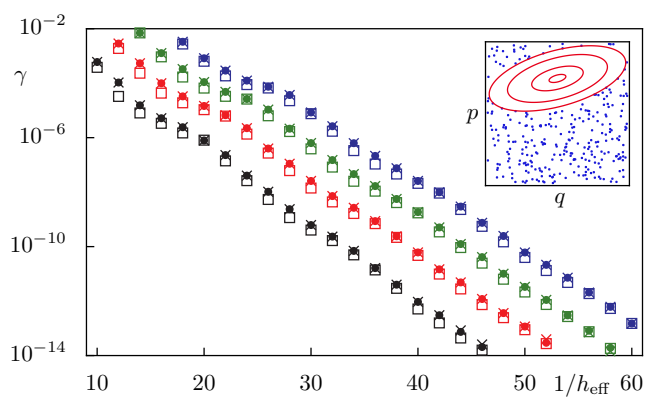}
    \caption{(Color online) Comparison of tunneling rates $\gamma$ 
             obtained by opening the system
             (dots), time evolution (crosses), and the evaluation of avoided 
             crossings (squares) vs $1/\heff$. 
             We use the map $\mapho$ (see Sec.~\ref{sec:maps:amphib:ho})
             and consider the regular states $m \leq 3$. 
             The inset shows the phase space of the system.
            }
    \label{fig:num:comparison}
  \end{center}
\end{figure}

To test the theoretical prediction derived in Sec.~\ref{sec:deriv}
we compare its results to numerical rates in Secs.~\ref{sec:maps:app} 
and \ref{sec:billiards}. 
In this section we present three alternative methods to numerically 
compute tunneling rates: 
(A) opening the system, (B) time evolution of regular states, and (C) evaluating  
avoided crossings. 
Figure~\ref{fig:num:comparison} shows a comparison of the tunneling rates 
obtained by these three methods for a quantum map. We find excellent agreement
between the first two methods while the last approach shows small deviations.

\subsection{Opening the system}
\label{sec:numerics:open}

The structure of the considered phase space, with one regular island 
surrounded by the chaotic sea, allows for the determination of tunneling rates 
by introducing absorption somewhere in the chaotic region of phase space. 
For quantum maps this can be realized, e.g., by using a non-unitary
open quantum map \cite{BorGuaShe1991,SchTwo2004}
\begin{eqnarray}
\label{eq:numerics:open_Uo}
 U^o = PUP,
\end{eqnarray}
where $P$ is a projection operator onto the complement of the absorbing region. 
An example is given by a sum of projectors on position eigenstates, 
\begin{eqnarray}
\label{eq:numerics:open_P}
P=\sum_{q_l}^{q_r} |q\rangle\langle q|, 
\end{eqnarray}
where the regular island is located well inside the interval $[q_l,q_r]$.

\begin{figure}[tb]
 \begin{center}
  \includegraphics[width=8.5cm]{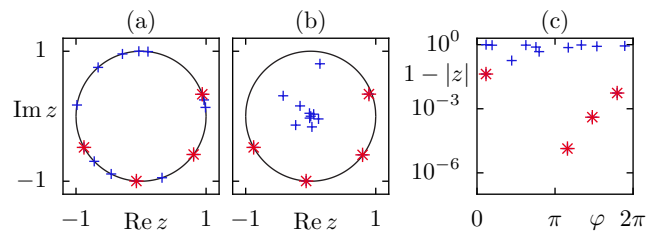}
  \caption[]
        {(Color online) (a) The eigenvalues of the unitary quantum map $\mapho$ 
         (see Sec.~\ref{sec:maps:amphib:ho}) with $\heff=1/14$ 
         are located on the unit circle.
         Regular states are marked by red stars, chaotic states by blue plus 
         symbols.
         (b) The eigenvalues of the open system $U^o$ are located inside the 
         unit circle. (c) The distance $1-|z|$ is shown on a logarithmic scale.
         }
         \label{fig:num:dist_unitcircle}
 \end{center}
\end{figure}

While the eigenvalues of $U$ are located on the unit circle the eigenvalues 
of $U^o$ are inside the unit circle as $U^o$ is sub-unitary, 
see Fig.~\ref{fig:num:dist_unitcircle}. The eigenequation of 
$U^o$ reads
\begin{eqnarray}
 U^o|\psi_{n}^{o}\rangle = z_n|\psi_{n}^{o}\rangle
\end{eqnarray}
with eigenvalues
\begin{eqnarray}
\label{eq:numerics:open_evals}
 z_n = \ue^{\ui\left(\varphi_n + \ui\frac{\gamma_n}{2} \right)}.
\end{eqnarray}
The decay rate is characterized by the imaginary part of the quasi-energies in
Eq.~\eqref{eq:numerics:open_evals} and one has
\begin{eqnarray}
\label{eq:numerics:open}
 \gamma_m = -2\log|z_m| \approx 2(1-|z_m|).
\end{eqnarray}

In order to obtain the open map $U^o$ practically,
we quantize the classical map on the cylinder 
$(q,p)\in[-\infty,\infty]\times[-1/2,1/2]$ with the periodically extended 
potential $V(q)$. This leads to an infinite dimensional unitary matrix $U$ 
in position representation \cite{ChaShi1986} and we find $U^o$ with 
Eq.~\eqref{eq:numerics:open_Uo} using the projector $P$ given by
Eq.~\eqref{eq:numerics:open_P}. After the diagonalization of $U^o$ we identify 
the eigenvalues $z_m$ of $U^o$ close to the unit circle, 
which correspond to the quasi-bound regular states, and use 
Eq.~\eqref{eq:numerics:open} to determine the tunneling rates, see 
Fig.~\ref{fig:num:comparison} (dots) for an example.

If the chaotic region does not contain partial barriers and shows no dynamical 
localization, it is justified to assume that the 
probability of escaping the regular island is equal to the probability of 
leaving through the absorbing regions located in the chaotic sea. 
Then, the location of the absorbing regions in the chaotic part of phase space
has no effect on the decay rates.

In generic systems, however, partial barriers will appear in the chaotic 
region of phase space. The additional transition through these structures 
further limits the quantum transport such that the calculated decay through 
the absorbing region occurs slower than the decay from the regular island to 
the neighboring chaotic sea. Similarly, dynamical localization in the chaotic
region may slow down the decay. The quantitative influence of partial barriers and 
dynamical localization on the regular-to-chaotic tunneling rates is an 
open problem for future studies. If necessary we will suppress their influence by 
moving the absorbing regions closer to the regular island.

\subsection{Time evolution}
\label{sec:numerics:time_evol}

\begin{figure}[tb]
  \begin{center}
    \includegraphics[width=8.5cm]{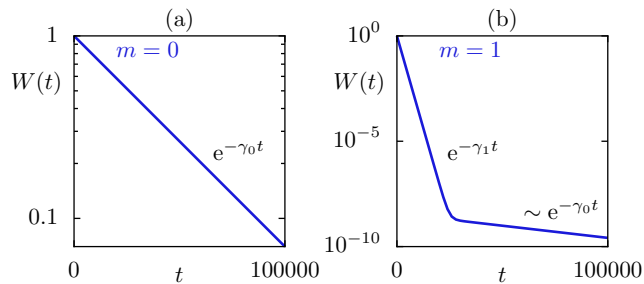}
    \caption{(Color online) Time evolution of a purely regular state for the system
             $\mapho$ (see Sec.~\ref{sec:maps:amphib:ho}) with $\heff=1/14$. 
             We present $W(t)$ vs $t$ obtained by 
             Eq.~\eqref{eq:maps:num:time_evol} on a semi-logarithmic
             scale. 
             (a) For the regular ground state the tunneling 
             rate to the chaotic sea $\gamma_0$ is determined by the slope 
             of the numerical data.
             (b) For the first excited state $m=1$ the slope for small times $t$
             determines the tunneling rate $\gamma_1$, 
             while for larger times the decay of the ground state $\gamma_0$ 
             with a smaller slope is found.
            }
    \label{fig:maps:time_evol}
  \end{center}
\end{figure}

A simple method to obtain a numerical prediction of the tunneling rates 
for quantum maps is given by the time-evolution of a purely 
regular state $|\psireg^m\rangle$ with a non-unitary operator 
$U^o = P U P$. Here $P$ projects onto a region in phase space which 
includes the regular island. We consider 
\begin{eqnarray}
\label{eq:maps:num:time_evol}
 W_m(t) = \Vert\Preg(U^o)^t|\psireg^m\rangle\Vert^2
\end{eqnarray}
for $t\in\N$, which describes the probability of the time-evolved regular state 
in the regular island at time $t$. 
At each time step some probability of $|\psireg^m\rangle$ is absorbed in the chaotic 
region due to the openness of the quantum map $U^o$. Consequently, $W_m(t)$ decays
exponentially, $W_m(t)\approx\ue^{-\gamma_m t}$, and the tunneling rates $\gamma_m$ can be 
determined by a fit of the numerical data. 
If $|\psireg^m\rangle$ contains admixtures from lower excited regular states
(with smaller tunneling rates) their decay dominates at times $t \gg 1/\gamma_m$.
If it contains admixtures from higher excited regular states (with larger 
tunneling rate) their decay will be seen at small times, 
see Fig.~\ref{fig:maps:time_evol}. The computed tunneling rates are in
excellent agreement with the results obtained by opening the system, 
see Fig.~\ref{fig:num:comparison} (crosses).
This method works best for regular states $|\psireg^m\rangle$ which 
resemble the corresponding eigenstates of the mixed system 
$U$ with high accuracy. It is particularly useful for Hilbert spaces of 
large dimension $N$ where diagonalizing the matrix $U^o$ would numerically
be very time-consuming.

\subsection{Evaluation of avoided crossings}
\label{sec:numerics:avcr}

\begin{figure}[tb]
 \begin{center}
  \includegraphics[width=8.5cm]{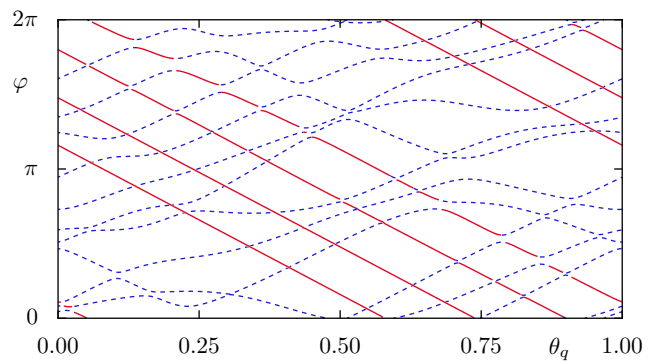}
  \caption[]
        {(Color online) Quasi-energies of the quantum map $U$ for the 
         system $\mapho$ (see Sec.~\ref{sec:maps:amphib:ho}) vs Bloch phase
         $\theta_q$ with $\heff=1/14$ and $\theta_p=0$. 
         Regular states (red solid lines) show avoided crossings with 
         chaotic eigenstates (blue dashed lines).}
         \label{fig:num:avcr}
 \end{center}
\end{figure}

The third method calculates the tunneling rate of a regular 
state directly from the spectrum of the system. 
For quantum maps we determine the quasi-energies $\varphi$ under variation of
a parameter of the system which leaves the classical dynamics invariant, such
as the Bloch phase $\theta_q$ or $\theta_p$. These phases specify the 
periodicity conditions on the torus and can be incorporated in the quantization
of the map \cite{KeaMezRob1999}.
Under variation of such a parameter the quasi-energy of the considered regular state
$\varphi_m$ shows avoided crossings with quasi-energies $\varphi_{\text{ch}}$
of chaotic states, see Fig.~\ref{fig:num:avcr}. 
These avoided crossings have widths $\Delta\varphi_{\text{ch},m}$.
According to degenerate perturbation theory they are related to the matrix 
elements $\vchm$ by $\Delta\varphi_{\text{ch},m}=2\vchm$ and fluctuate 
depending on the involved chaotic state. The tunneling rate follows 
from the dimensionless version of Fermi's golden rule, Eq.~\eqref{eq:app:fgr:2},
\begin{eqnarray}
\label{eq:numerics:avcr_maps}
 \gamma_m & = & \frac{N_{\text{ch}}}{4} \langle|\Delta\varphi_{\text{ch},m}|^2
                \rangle_{\text{ch}},
\end{eqnarray}
where $N_{\text{ch}}$ is the number of chaotic states.
Note, that the two methods discussed in 
Secs.~\ref{sec:numerics:open} and \ref{sec:numerics:time_evol}
determine the tunneling rates $\gamma$ for fixed Bloch phases.
Under variation of $\theta_q$ or $\theta_p$ we observe numerically for these
methods that the rates vary by at most a factor of $2$, while
Eq.~\eqref{eq:numerics:avcr_maps} gives an average rate.

The quality of this prediction depends on the number of 
avoided crossings entering the average in Eq.~\eqref{eq:numerics:avcr_maps}.
If only a few avoided crossings are included, the statistical error of the
result is large. 
Note, that for quantum maps with a mean drift in the chaotic sea,
such as the systems $\mapd$ (introduced in Sec.~\ref{sec:maps:amphib:ho}), 
under variation of the Bloch phases $\theta$ 
several chaotic states will show avoided crossings with each regular state
\cite{BerBalTabVor1979}.
The results of this method are presented in Fig.~\ref{fig:num:comparison} 
(squares) for an example system. 
We typically obtain tunneling rates
which are smaller than the results of the other two methods. While we have no 
explanation of this behavior, the deviation is smaller than a factor of two, which is 
sufficient for a comparison to theoretical predictions.

Also for billiards tunneling rates can be computed by this method.
We determine the spectrum under variation of parts of the billiard 
boundary which leaves the classically regular dynamics unchanged but 
affects the chaotic dynamics. Quantum mechanically, the eigenenergies of the 
regular states remain almost unaffected while the eigenenergies of 
the chaotic states vary strongly, due to the changing density of chaotic 
states. Hence, avoided crossings of widths 
$\Delta E_{\text{ch},mn}$ between the considered regular 
and the chaotic states appear. 
The tunneling rate is given by Fermi's golden rule, Eq.~\eqref{eq:deriv:FGR_bil},
\begin{equation}
\label{eq:numerics:avcr}
 \Gamma_{mn} = \frac{2\pi}{\hbar} \frac{\langle 
              |\Delta E_{\text{ch},mn}|^2 \rho_{\text{ch}}\rangle}{4} \approx 
 \frac{\hbar}{16M} \langle |\Delta E_{\text{ch},mn}|^2 \Achb\rangle,  
\end{equation}
where we average the product of all numerically determined widths 
$\Delta E_{\text{ch},mn}=2\Vchmn$ and the corresponding density of chaotic 
states $\rho_{\text{ch}}$, which we approximate by 
its leading Weyl term, see Sec.~\ref{sec:deriv:billiard}.
Note, that in general it can be difficult 
to deform a part of the billiard boundary such that the regular dynamics is 
unchanged while still the numerical methods for the determination of eigenvalues 
in billiard systems are applicable.


\section{Application to quantum maps}
\label{sec:maps:app}

In the following we will apply the fictitious integrable system approach, 
derived in Sec.~\ref{sec:deriv}, to the prediction of
direct regular-to-chaotic tunneling rates in the case of quantum 
maps and compare the results to numerical rates for different example systems.

\subsection{Designed maps $\mapd$}
\label{sec:maps:amphib}

Our aim is to introduce kicked systems which can be designed such that
their phase space shows one regular island embedded in the chaotic sea, with 
very small nonlinear resonance chains within the regular island, a negligible 
hierarchical region, and without relevant partial barriers in the chaotic component.
For such a system it is possible to study the direct regular-to-chaotic
tunneling process without additional effects caused by these structures.

To this end we define the family of maps $\mapd$, according to 
Eq.~\eqref{eq:maps:kicked_map},
with an appropriate choice of the functions $T'(p)$ and $V'(q)$ 
\cite{ShuIke1995,SchOttKetDit2001,BaeKetMon2005,BaeKetLoeSch2008,LoeBaeKetSch2010}.
For this we first introduce
\begin{eqnarray}
\label{sec:maps:amphib:tv_disc}
 t'(p) & = & \left\{ \begin{array}{llrcl}
                   \tfrac{1}{2} - (1-2p) & \text{ for } & -\tfrac{1}{2} < & p & < 0 \\
                   \tfrac{1}{2} + (1-2p) & \text{ for } & \,0 < & p & < \tfrac{1}{2},
                     \end{array} \right. \\
 v'(q) & = &  \begin{array}{llrcl}
                   -rq+Rq^2 & \hspace{0.51cm}\text{ for } 
                    &-\tfrac{1}{2} < & q & < \tfrac{1}{2} 
                     \end{array}
\end{eqnarray}
with $0<r<2$ and $R \geq 0$. This gives a regular island around $(q,p)=(0,1/4)$.
Considering periodic boundary conditions the functions $t'(p)$ 
and $v'(q)$ show discontinuities at $p=0,\pm 1/2$ and
$q=\pm 1/2$, respectively. In order to avoid these discontinuities we smooth 
the periodically extended functions $v'(q)$ and $t'(q)$ with a Gaussian, 
\begin{equation}
\label{sec:maps:amphib:gaussian}
  G(z)=\frac{1}{\sqrt{2\pi\varepsilon^2}}\exp\left(-\frac{z^2}{2\varepsilon^2} \right),
\end{equation}
resulting in analytic functions 
\begin{eqnarray}
\label{sec:maps:amphib:tv_cont}
 T'(p) & = & \Int \ud z \, t'(z)G(p-z),\\  
 V'(q) & = & \Int \ud z \, v'(z)G(q-z),
\end{eqnarray}
which are periodic with respect to the phase-space unit cell. With this we 
obtain the maps $\mapd$ depending on the 
parameters $r$, $R$, and the smoothing strength $\varepsilon$. 
The smoothing $\varepsilon$ determines the size of the hierarchical region
at the border of the regular island. Tuning the parameters 
$r$ and $R$ one can find situations, 
where all nonlinear resonance chains inside the regular island are small.

\subsubsection{Map $\mapho$ with harmonic oscillator-like island}
\label{sec:maps:amphib:ho}

\begin{figure}[tbh!]
 \begin{center}
  \includegraphics[width=8.5cm]{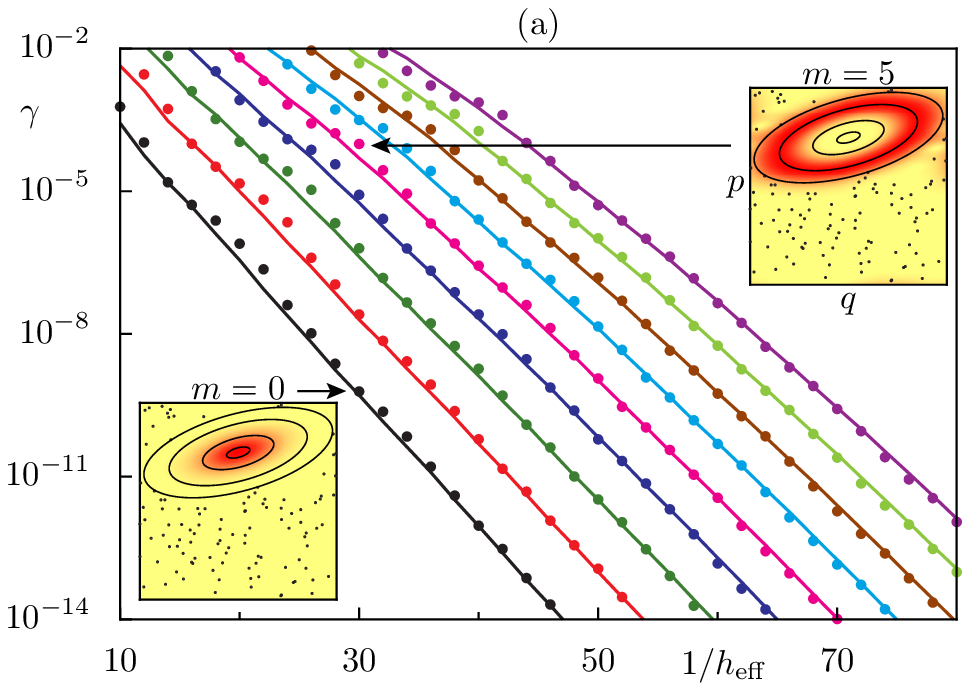}\\
  \includegraphics[width=8.5cm]{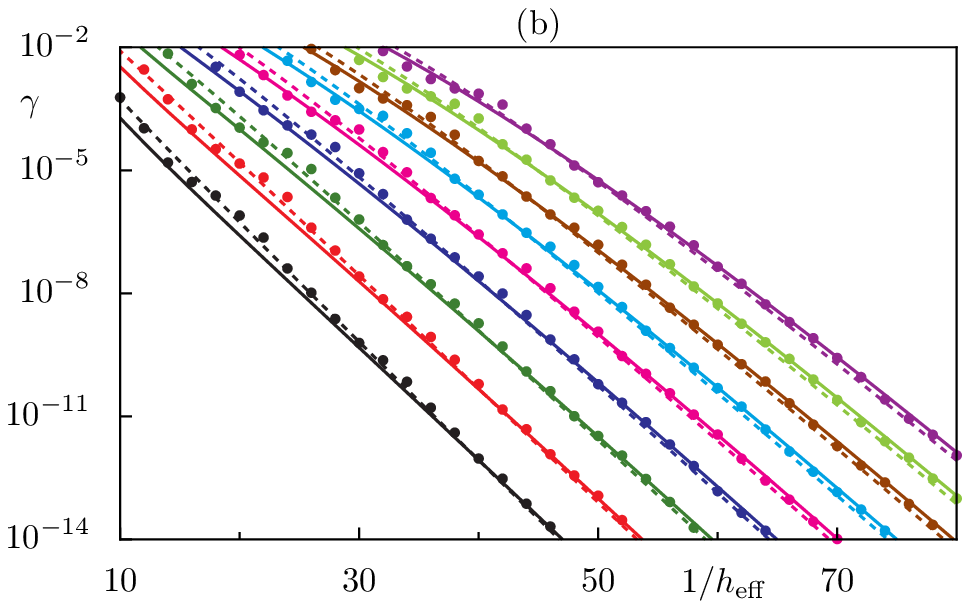}\\
  \includegraphics[width=8.5cm]{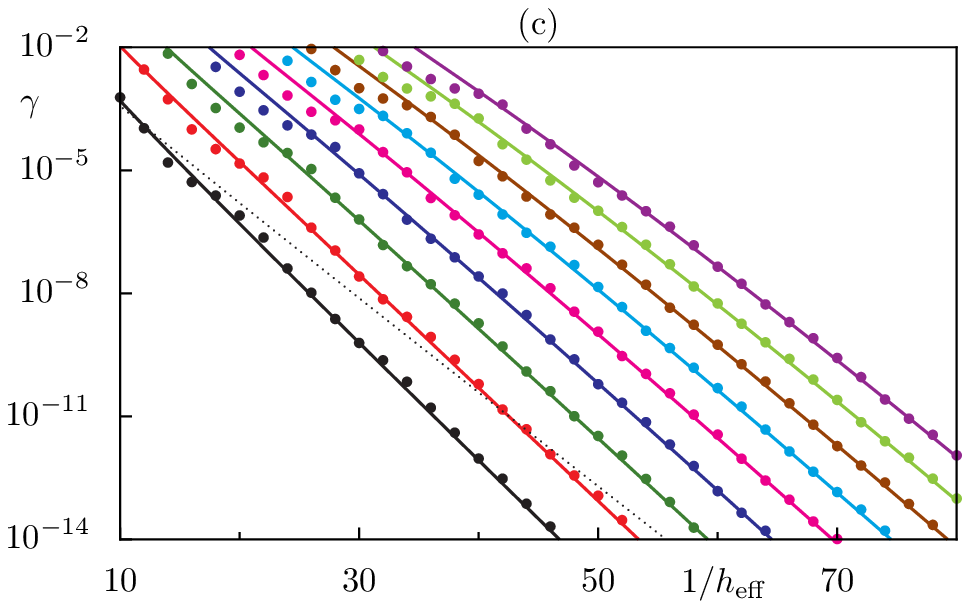}
  \caption[]
        {(Color online) Numerical tunneling rates (dots) for $m \leq 8$ 
         for the map $\mapho$ with a harmonic oscillator-like island.
         (a) Comparison with Eq.~\eqref{eq:deriv:final_result} (lines). 
         The insets show Husimi functions of the regular states for the 
         quantum numbers $m=0$ and $m=5$ at $1/\heff=30$ and the classical phase 
         space of the system.
         (b) Comparison with Eqs.~\eqref{eq:maps:sc:tunneling_rate_2} (solid lines)
         and \eqref{eq:maps:sc:tunneling_rate_2b} (dashed lines).
         (c) Comparison with Eq.~\eqref{eq:maps:sc:wkb_formula} (solid lines).
         The prediction of Refs.~\cite{PodNar2003,She2005}, 
         Eq.~\eqref{eq:maps:amphib:PodNar}, 
         for $m=0$ with a fitted prefactor is shown (dotted line).}
         \label{fig:maps:rates_amph_ellipt}
 \end{center}
\end{figure}

For $R=0$ both functions $v'(q)$ and $t'(p)$ are linear in $q$ and $p$,
respectively. In this case we find a harmonic oscillator-like regular 
island with elliptic invariant tori and constant rotation number. 
We choose the parameters $r=0.46$, $R=0$, $\varepsilon=0.005$ and 
label the resulting map by $\mapho$. Its phase-space 
is shown in the insets of Fig.~\ref{fig:maps:rates_amph_ellipt}. 
Numerically, we determine tunneling rates by introducing absorbing regions 
at $|q| \geq 1/2$, as described in Sec.~\ref{sec:numerics:open}.
In order to apply the fictitious integrable system approach
we use the Hamiltonian of a harmonic oscillator as 
$\Hreg$. It is squeezed and tilted according to the linearized dynamics in the
vicinity of the stable fixed point located at the center of the regular island. 
Its eigenfunctions $|\psireg^m\rangle$ are analytically known,
see Appendix~\ref{sec:appendix:ho_states}.

Figure~\ref{fig:maps:rates_amph_ellipt}(a) shows the numerically evaluated 
prediction of Eq.~\eqref{eq:deriv:final_result} compared to numerical tunneling rates. 
We find excellent agreement over more than $10$ orders of magnitude in $\gamma$.
In the regime of large tunneling rates small deviations occur which can 
be attributed to the influence of the chaotic sea on the regular states:
These states are located on quantizing tori close to the border of 
the regular island and are affected by the regular-to-chaotic transition region. 
However, the deviations in this regime are smaller than a factor of two.

Figure~\ref{fig:maps:rates_amph_ellipt}(b) shows the results of 
Eq.~\eqref{eq:maps:sc:tunneling_rate_2},  
which are obtained by approximating $\Ureg$
by a kicked system, $\Uregt=\UVreg U_T$, again using the analytically given
$|\psireg^m\rangle$. These results are still in excellent agreement with the 
numerical rates (solid lines). In Eq.~\eqref{eq:maps:sc:tunneling_rate_2b} the 
sum over the positions is replaced by an integral, which explains the 
small deviations to the results of Eq.~\eqref{eq:maps:sc:tunneling_rate_2}, see
Fig.~\ref{fig:maps:rates_amph_ellipt}(b) (dashed lines). 
These deviations vanish in the semiclassical limit. 

Finally, in Fig.~\ref{fig:maps:rates_amph_ellipt}(c) we compare the results of 
the semiclassical prediction, Eq.~\eqref{eq:maps:sc:wkb_formula}, to the numerical rates.
Due to the approximations performed in the derivation of this formula
stronger deviations are visible in the regime of large tunneling rates
while the agreement in the semiclassical regime is still excellent.

In Refs.~\cite{PodNar2003,She2005} a prediction was derived 
for the tunneling rate of the regular ground state,
\begin{eqnarray}
\label{eq:maps:amphib:PodNar}
 \gamma_0 = c \frac{\Gamma(\alpha,4\alpha)}{\Gamma(\alpha,0)},
\end{eqnarray}
where $\Gamma$ is the incomplete gamma function, $\alpha=\Areg/\heff$,
and $c$ is a constant.
Equation~\eqref{eq:maps:amphib:PodNar} can be approximated semiclassically 
\cite{SchEltUll2005}, $\alpha\to\infty$, leading to
\begin{eqnarray}
\label{eq:maps:amphib:PodNar2}
 \gamma_0 \propto \frac{1}{\sqrt{\alpha}}\ue^{-\alpha (3-\ln 4)}.
\end{eqnarray}
Figure~\ref{fig:maps:rates_amph_ellipt}(c) shows the comparison
of Eq.~\eqref{eq:maps:amphib:PodNar} (dotted line) to the numerical 
rates for the map $\mapho$. Especially
in the semiclassical regime strong deviations are visible. The factor
$2$ which appears in the exponent of Eq.~\eqref{eq:maps:sc:wkb_formula}
is more accurate than the factor $3-\ln 4$ in 
Eq.~\eqref{eq:maps:amphib:PodNar2}.

\subsubsection{Map $\mapdef$ with deformed island}
\label{sec:maps:amphib:def}

\begin{figure}[tb]
  \begin{center}
    \includegraphics[width=8.5cm]{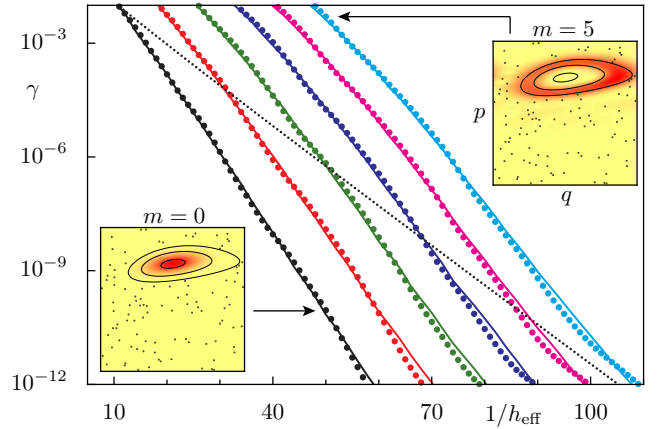}
    \caption{(Color online) Dynamical tunneling rates from a regular island 
          to the chaotic sea for the map $\mapdef$. 
          Numerical rates (dots) and prediction 
          following from Eq.~\eqref{eq:deriv:final_result} (lines) 
          vs $1/\hbareff$ for quantum numbers $m \leq 5$.
          The insets show Husimi representations of the regular states 
          $m=0$ and $m=5$ at $1/\heff=50$. The prediction of 
          Refs.~\cite{PodNar2003,She2005}, Eq.~\eqref{eq:maps:amphib:PodNar}, 
          for $m=0$ with a fitted prefactor is shown (dotted line).
         }
    \label{fig:maps:rates_amph_dist}
  \end{center}
\end{figure}

In generic systems the regular island has a non-elliptic shape and the 
rotation number of regular tori changes from the center of the 
regular region to its border with the chaotic sea. Such a situation
can be achieved for the family of maps $\mapd$ with the parameter $R\neq 0$.
For most combinations of the parameters $r$ and $R$ resonance structures
appear inside the regular island. They limit the $\heff$-regime in which the
direct regular-to-chaotic tunneling process dominates. Hence, we choose a 
situation in which the nonlinear resonances are small such that
their influence on the tunneling process is expected only at large $1/\heff$.
For this we use $r=0.26$, $R=0.4$, $\varepsilon=0.005$ and label the resulting 
map with a deformed island by $\mapdef$, see the inset 
in Fig.~\ref{fig:maps:rates_amph_dist} for its phase space.

We determine the fictitious integrable system $\Hreg$ by means of the Lie-transformation 
method described in Sec.~\ref{sec:regsyst}. It is then quantized and its 
eigenfunctions are determined numerically. 
Figure~\ref{fig:maps:rates_amph_dist} shows a comparison of the numerically 
evaluated prediction of Eq.~\eqref{eq:deriv:final_result} (solid lines) to
numerical tunneling rates (dots) yielding excellent agreement for
$\gamma\gtrsim 10^{-11}$. For smaller values of $\gamma$ 
deviations occur due to resonance-assisted tunneling which is caused by a 
small $10$:$1$ resonance chain. 
Similar to the case of the harmonic oscillator-like island the fictitious 
integrable system $\Ureg$ can be approximated by a kicked
system $\Ureg\approx\UVreg\UT$ using $\widetilde{V}(q)=-rq^2/2+Rq^3/3$. 
Hence, Eqs.~\eqref{eq:maps:sc:tunneling_rate_2} and 
\eqref{eq:maps:sc:tunneling_rate_2b} can be evaluated giving similarly 
good agreement (not shown).
The prediction of Eq.~\eqref{eq:maps:amphib:PodNar} \cite{PodNar2003,She2005} 
(dotted line) shows large deviations to the numerical rates.

\subsubsection{Map $\mapwc$ with weakly chaotic dynamics}
\label{sec:maps:amphib:wc}

In Ref.~\cite{SheFisGuaReb2006} the dynamical tunneling 
process from one regular island to the chaotic sea in a system 
with weakly chaotic dynamics was investigated. 
Here, tunneling can occur to regions in phase space which are
far away from the border of the regular island with the chaotic sea. 
We show that also in this situation, which we believe to be non-generic, 
the fictitious integrable system approach can be applied. 
Its results will be compared to the WKB prediction of 
Ref.~\cite{SheFisGuaReb2006}.

\begin{figure}[tb]
  \begin{center}  
    \includegraphics[width=8.5cm]{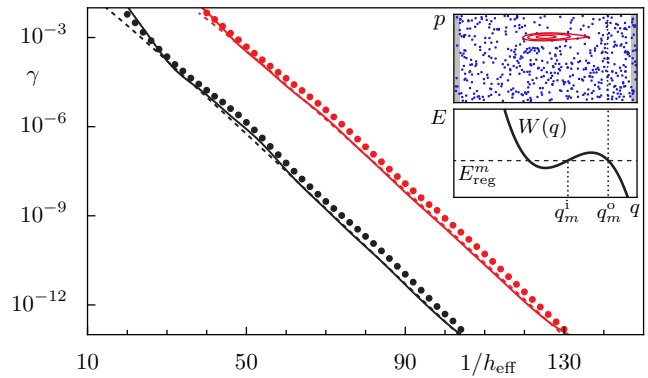}
    \caption{(Color online) Dynamical tunneling rates from a 
             regular island to the chaotic sea 
             for the weakly chaotic system $\mapwc$.
             We compare numerical rates (dots) and the 
             prediction following from Eq.~\eqref{eq:deriv:final_result}
             (solid lines) and Eq.~\eqref{eq:maps:amphib:gamma_near_int} 
             \cite{SheFisGuaReb2006} (dashed lines) vs $1/\heff$ for 
             $m=0$ and $m=1$.             
             The upper inset shows the phase space of the system, where the
             absorbing regions are indicated in gray.  
             In the lower inset the approximate one-dimensional
             potential $W(q)$ used in Eq.~\eqref{eq:maps:amphib:gamma_near_int} 
             is presented with the inner (outer) turning point 
             $q_{m}^{\text{i}}$ ($q_{m}^{\text{o}}$) at energy $\Ereg^m$.
            }
    \label{fig:maps:near_integrable:gamma}
  \end{center}
\end{figure}

A system with weakly chaotic dynamics can be modeled by the example 
systems $\mapd$. We choose $r=0.05$, $R=0.1$, and $\varepsilon=0.005$,
consider the extended phase space $(q,p)\in[-1,1]\times[-1/2,1/2]$, 
and label the resulting map by $\mapwc$. Its phase space is shown in the 
upper inset of Fig.~\ref{fig:maps:near_integrable:gamma}. 

In order to apply the fictitious integrable system approach
we use the Lie-transformation method, as described in Sec.~\ref{sec:regsyst},
to obtain the fictitious integrable system $\Hreg$.
As the system $\mapwc$ is only weakly driven, due to the small 
parameters $r$ and $R$, it is sufficient to consider the zeroth order of
the Lie expansion which has no mixed terms containing $q$ and $p$ 
simultaneously. The resulting integrable approximation
\begin{eqnarray}
\label{eq:maps:amphib:hamilt_near_int}
 \Hreg(q,p) = \frac{p^2}{2}+W(q),
\end{eqnarray}
describes the dynamics in a potential $W(q)=\omega^2q^2/2-Rq^3/6$ with 
$\omega=\sqrt{r/2}$, see the lower inset in 
Fig.~\ref{fig:maps:near_integrable:gamma}. 

In Ref.~\cite{SheFisGuaReb2006} it was shown that for such a weakly chaotic 
system regular-to-chaotic tunneling rates can be predicted by 
one-dimensional tunneling under the energy barrier of the potential 
$W(q)$. For the tunneling rates one finds 
\begin{equation}
\label{eq:maps:amphib:gamma_near_int}
 \gamma_m \approx \frac{\omega_m}{2\pi} \exp\left(-\frac{2}{\hbareff}
 \Int_{q_{m}^{\text{i}}}^{q_{m}^{\text{o}}}|p(q,\Ereg^m)|\,\ud q\right)
\end{equation}
where $p(q,\Ereg^m)= \sqrt{2\Ereg^m-\omega^2q^2+Rq^3/3}$, 
$q_{m}^{\text{i}}$ denotes the right classical turning point inside the 
potential well, $q_{m}^{\text{o}}$ is the turning point outside the 
potential well, and $\omega_m$ is the oscillation period on the 
$m$th quantizing torus. The eigenenergies $\Ereg^m$
can be calculated using the Bohr-Sommerfeld quantization
$\oint p(q,\Ereg^m)\,\ud q = \heff(m+1/2)$.
For the system $\mapwc$ the right turning point $q_{m}^{\text{o}}$
is located far away from the regular island. Tunneling occurs
to the region with $q>q_{m}^{\text{o}}$ deep inside the weakly chaotic sea and not to
the neighborhood of the regular island, as for the other examples considered in 
this paper. 

In Fig.~\ref{fig:maps:near_integrable:gamma} we compare the numerically 
evaluated prediction of Eq.~\eqref{eq:deriv:final_result} (solid lines) to
the result of Eq.~\eqref{eq:maps:amphib:gamma_near_int} (dashed lines) and 
numerical rates (dots), which are determined by absorbing regions 
at $|q| \geq 1$. We find good agreement. 

Note, that for the weakly chaotic system $\mapwc$ the purely regular 
states $|\psireg^m\rangle$ of $\Hreg$ show the correct tunneling tails 
far beyond the regular island including the outer turning point $q_{m}^{\text{o}}$.
Moreover, the semiclassical evaluation of Eq.~\eqref{eq:deriv:final_result_app}, 
presented in Sec.~\ref{sec:maps:sc}, can be performed. 
This leads to Eq.~\eqref{eq:maps:sc:tunnelrate_rand_wkb_einges} 
which has the same exponential term as 
Eq.~\eqref{eq:maps:amphib:gamma_near_int} but a different prefactor. 

We want to emphasize that generically regular-to-chaotic tunneling cannot be
described by Eq.~\eqref{eq:maps:amphib:gamma_near_int}, as the integrable 
approximation $\Hreg$ is not of the form 
Eq.~\eqref{eq:maps:amphib:hamilt_near_int}.

\subsection{Standard map}
\label{sec:maps:stmap}

\begin{figure}[tb]
  \begin{center}
    \includegraphics[width=8.5cm]{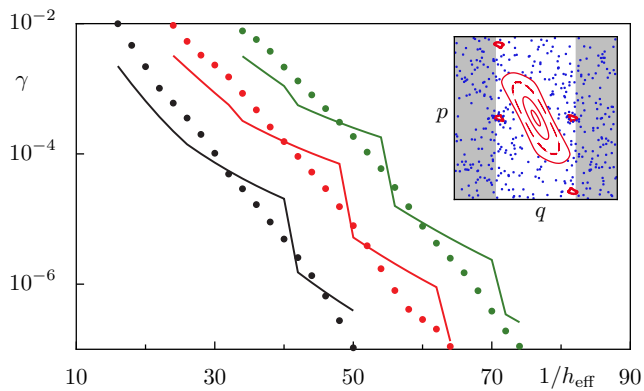}
    \caption{(Color online) Tunneling rates for the standard map ($\kappa=2.9$) 
          for $m \leq 2$ vs $1/\heff$. 
          Prediction of Eq.~\eqref{eq:deriv:final_result} 
          (lines) and numerical rates (dots), obtained using absorbing 
          regions at $|q| \geq 1/4$ (gray-shaded area of the inset). 
          }
    \label{fig:maps:rates_stmap}
  \end{center}
\end{figure}

The paradigmatic model of an area preserving map is the standard map 
\cite{Chi1979}, defined by Eq.~\eqref{eq:maps:kicked_map} with the functions
\begin{eqnarray}
\label{sec:maps:stmap:TV}
 T'(p) & = & p, \\
 V'(q) & = & \frac{\kappa}{2\pi}\sin(2\pi q).
\end{eqnarray}
For $\kappa$ between $2.5$ and $3.0$ one has a large generic 
regular island with a relatively small hierarchical region surrounded by a
$4$:$1$ resonance chain, see the inset in Fig.~\ref{fig:maps:rates_stmap}. 

When determining tunneling rates numerically by introducing absorbing regions at 
$|q| \geq 1/2$ we find strong fluctuations as a function of  
$\heff$, presumably caused by partial barriers. 
Using absorption at $|q| \geq 1/4$, which is closer to the island, we find 
smoothly decaying tunneling rates (dots in Fig.~\ref{fig:maps:rates_stmap}).
 
Evaluating Eq.~\eqref{eq:deriv:final_result} for $\kappa=2.9$ 
gives reasonable agreement with these numerical rates with deviations
up to a factor of $5$, 
see Fig.~\ref{fig:maps:rates_stmap} (solid lines).  
Here we determine $\Hreg$ using the method based on the frequency 
map analysis as the Lie transformation is not able to reproduce
the dynamics within the regular island of $U$, see Sec.~\ref{sec:regsyst}.  
With increasing order $K$ of the expansion series of $\Hreg$ the
tunneling rates following from Eq.~\eqref{eq:deriv:final_result} diverge, 
see Fig.~\ref{fig:maps:convergence}(b). 
Hence, for the predictions in Fig.~\ref{fig:maps:rates_stmap} 
we choose $K=2$ which is the largest order before the divergence starts to set in.
Note, that at such small order $K$ the accuracy of $\Hreg$ within the regular 
region of $U$ is inferior compared to the examples discussed before. 
Hence, in Eq.~\eqref{eq:deriv:final_result} the state $U|\psireg^m\rangle$
has small contributions of other purely regular states $|\psireg^n\rangle$
in the regular island. 
These contributions are compensated by the application of the projector 
$\Pch$. However, this projector depends on the number of regular states 
$\mmax$, which grows in the semiclassical limit. 
If $\mmax$ increases by one, $\Preg$ suddenly projects onto a larger
region in phase space. This explains the steps of the theoretical prediction, 
Eq.~\eqref{eq:deriv:final_result}, visible in Fig.~\ref{fig:maps:rates_stmap}.
How to improve $\Hreg$ and the projector $\Pch$ is an open question.

\subsection{Map $\mapsi$ with a regular stripe}
\label{sec:maps:shudo}

\begin{figure}[tb]
  \begin{center}
    \includegraphics[width=8.5cm]{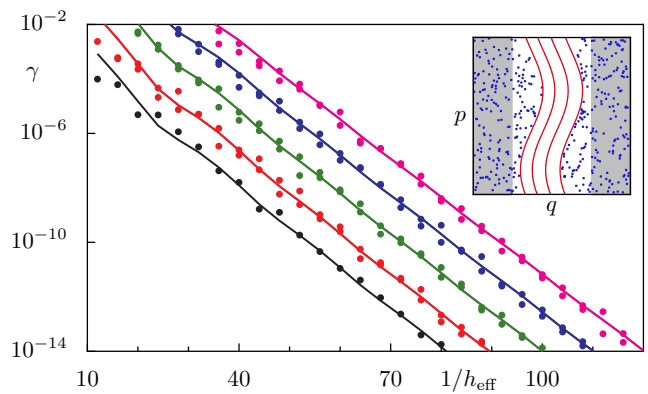}
    \caption{(Color online) Dynamical tunneling rates from a regular stripe 
             to the chaotic sea for the map $\mapsi$. 
             We compare numerical rates (dots) and the 
             prediction following from Eq.~\eqref{eq:deriv:final_result}
             (lines) vs $1/\heff$ for the quantum numbers $|m| \leq 4$. 
             The inset shows the phase space of the system. The numerical rates
             are obtained using absorbing regions at $|q| \geq 1/4$ 
             (gray-shaded area of the inset).
          }
    \label{fig:maps:rates_shudo_map}
  \end{center}
\end{figure}

Another designed kicked system was introduced in 
Refs.~\cite{ShuIke1995,ShuIke1998,IshTanShu2007,IshTanShu2009}. 
Here the regular region consists of a stripe in phase space, see the inset in
Fig.~\ref{fig:maps:rates_shudo_map}. 
In our notation the mapping $\mapsi$, Eq.~\eqref{eq:maps:kicked_map}, 
is specified by the functions 
\begin{eqnarray}
 \label{eq:maps:shudo:v}
 V'(q) & = & -\frac{1}{2\pi}(8\pi aq+d_1-d_2\nonumber\\
       &   &  +\frac{1}{2}[8\pi aq-\omega+d_1] \tanh[b(8\pi q-q_d)]\nonumber \\
       &   & +\frac{1}{2}[-8\pi aq+\omega+d_2]\tanh[b(8\pi q+q_d)]),\;\;\;\\
 \label{eq:maps:shudo:t}
 T'(p) & = & -\frac{K}{8\pi}\sin(2\pi p)
\end{eqnarray}
with parameters $a=5$, $b=100$, $d_1=-24$, $d_2=-26$, $\omega=1$, $q_d=5$, and $K=3$. 
The kinetic energy $T(p)$ is periodic with respect to the phase space unit cell. 

The resulting map $\mapsi$ is similar to the system $\mapho$
as it also destroys the integrable region by smoothly 
changing the function $V'(q)$ at $|q|=q_d/(8\pi)$. 
For $|q| < q_d/(8\pi)$ the potential term is almost linear while it tends to 
the standard map for $|q| > q_d/(8\pi)$. The parameter $b$ determines the width
of the transition region.
In Ref.~\cite{IshTanShu2007} this map is used to study the evolution of a wave 
packet initially started in the regular region by means of complex paths. 
We now predict direct regular-to-chaotic tunneling rates with 
Eq.~\eqref{eq:deriv:final_result}. 
The fictitious integrable system $\Ureg$ is determined by continuing the dynamics 
within $|q|<q_d/(8\pi)$ to the whole phase space. It is given as a kicked system, 
Eq.~\eqref{eq:maps:kicked_map}, defined by the functions
\begin{eqnarray}
 \label{eq:maps:shudo:tvreg}
 \widetilde{V}'(q) & = & -\frac{1}{2\pi}
                         \left(\omega+\frac{d_1}{2}-\frac{d_2}{2}\right),\\
 T'(p) & = & -\frac{K}{8\pi}\sin(2\pi p).
\end{eqnarray}

When determining tunneling rates numerically using absorbing regions at 
$|q| \geq 1/2$ we find strong fluctuations as a function of  
$\heff$, similar to the standard map. 
Choosing $|q| \geq 1/4$ for the opening, which is closer 
to the regular stripe, we find smoothly decaying tunneling rates (dots in
Fig.~\ref{fig:maps:rates_shudo_map}). Their comparison with the numerical 
evaluated prediction of Eq.~\eqref{eq:deriv:final_result} shows 
excellent agreement, see Fig.~\ref{fig:maps:rates_shudo_map} (lines).

Note, that due to the symmetry of the map there are always two regular
states with comparable tunneling rates except for the ground state $m=0$.
These two states are located symmetrically around the center of the regular 
stripe. While the prediction,
Eq.~\eqref{eq:deriv:final_result}, is identical for both of these states, the
numerical results differ slightly due to the different chaotic dynamics in the
vicinity of the left and right borders of the regular region.


\section{Application to billiards}
\label{sec:billiards}

Billiards play a central role in both experimental and theoretical studies 
in quantum chaos. 
They are dynamical systems given by a point particle of mass $M$ which 
moves with constant velocity inside a domain $\Omega\in\R^2$ which we assume
to be compact. 
The particle is elastically reflected at the boundary $\partial\Omega$ 
such that the angle of incidence equals the angle of reflection. 
While there are only a few integrable and completely chaotic billiards, 
the majority shows a mixed phase space consisting of regions of regular and 
chaotic dynamics. 
Quantum mechanically, billiards are described by the time-independent Schr\"odinger
equation (in units $\hbar=2M=1$ used in this section)
\begin{eqnarray}
\label{eq:billiards:sgl}
 -\Delta \psi_n(\bq) = E_n\psi_n(\bq),\quad \bq\in\Omega
\end{eqnarray}
with the Dirichlet boundary condition $\psi_n(\bq)=0$, $\bq\in\partial\Omega$.
In Eq.~\eqref{eq:billiards:sgl} $\Delta$ denotes the Laplace operator
in two dimensions.
Equation~\eqref{eq:billiards:sgl} is identical to the eigenvalue
problem of the two-dimensional Helmholtz equation which for example describes 
electro-magnetic modes in a microwave cavity. This equivalence
allows for the simulation of quantum billiards by experiments using 
microwave cavities \cite{StoSte1990,Sri1991,GraHarLenLewRanRicSchWei1992,
SteSto1992,AltBaeDemGraHofRehRic1998}. 
   
The state of a particle is described by a wave function 
$\psi(\bq) \in L^2(\Omega)$ in position representation, where $L^2(\Omega)$ 
is the Hilbert space of square integrable functions on $\Omega$. Due to the 
compactness of $\Omega$ the eigenvalues $\{E_n\}$ are discrete and can be 
ordered as $0 \leq E_1 \leq E_2 \leq E_3 \leq \cdots$. The eigenfunctions can be 
chosen real and form an orthonormal basis on $L^2(\Omega)$.
In contrast to the case of quantum maps discussed in 
Sec.~\ref{sec:maps:app} one gets infinitely many  
eigenvalues and eigenfunctions.
There are only a few billiard systems, whose eigenfunctions are analytically 
known, e.g., the rectangular, circular, and elliptical billiard.
Usually an analytical solution of Eq.~\eqref{eq:billiards:sgl} 
is not possible.

The determination of tunneling rates for two-dimensional billiard systems is 
of current interest. It is relevant, e.g., in the 
context of light emission in optical microcavities 
\cite{WieHen2006,WieHen2008,ShiHarFukHenSasNar2010,YanLeeMooLeeKimDaoLeeAn2010}, 
flooding of regular
states \cite{BaeKetMon2005,BaeKetMon2007,Bit2010} and conductance properties of 
electrons in disordered wires with a magnetic 
field \cite{FeiBaeKetRotHucBur2006,FeiBaeKetBurRot2009}. 
Previous theoretical predictions of tunneling rates 
\cite{BarBet2007} or energy-splittings \cite{DorFri1995,FriDor1998} in 
billiards required additional free parameters.
 
We apply the fictitious integrable system approach in order to determine 
direct regular-to-chaotic tunneling rates for billiards. For this we employ 
Eqs.~\eqref{eq:deriv:coupmatelorth_bil} and \eqref{eq:deriv:FGR_bil}, where 
in the following we omit the tilde of the non-orthogonal chaotic states
$\psicht$ and label the corresponding dimensionless 
matrix elements and tunneling rates by $\vchmn$ and $\gamma_{mn}$, respectively. 
For the chaotic states entering in
Eq.~\eqref{eq:deriv:coupmatelorth_bil} we employ random 
wave models \cite{Ber1977} such that the average in Eq.~\eqref{eq:deriv:FGR_bil} 
becomes an ensemble average over the different realizations of the random wave 
model.
From this we obtain explicit analytical predictions for the mushroom 
billiard \cite{BaeKetLoeRobVidHoeKuhSto2008}, the annular billiard, 
two-dimensional nanowires with one-sided surface disorder, and 
optical microcavities \cite{BaeKetLoeWieHen2009}.

\subsection{Mushroom billiard}
\label{sec:billiards:mushroom}

\begin{figure}[tb]
  \begin{center}
    \includegraphics[width=6.6cm]{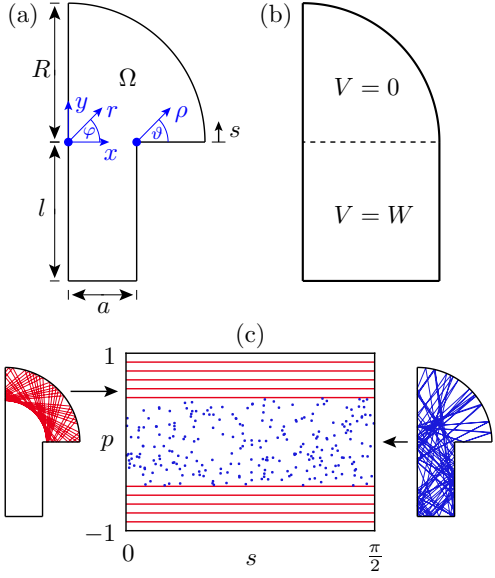}
    \caption{(Color online) (a) Schematic picture of the mushroom billiard, 
             with cap radius $R$, stem width $a$, and stem height $l$,
             showing the two coordinate systems used in the theoretical derivation
             of the direct regular-to-chaotic tunneling rates.
             (b) Auxiliary billiard $\Hreg^W$.
             (c) Poincar\'e section at the quarter-circle boundary 
             (relative tangential momentum $p$ vs arclength $s$) 
             showing regular and chaotic regions with illustrations of 
             trajectories.
            }
    \label{fig:billiards:mushroom:general}
  \end{center}
\end{figure}

We consider the desymmetrized mushroom billiard \cite{Bun2001}, 
see Fig.~\ref{fig:billiards:mushroom:general}(a), 
characterized by the radius $R$ of the quarter
circular cap, the stem width $a$, and the stem height $l$. 
This billiard is of great current interest 
\cite{VidStoRobKuhHoeGro2007,BarBet2007,AltMotKan2005,TanShu2006,DieFriMisRicSch2007} 
due to its sharply separated regular and chaotic regions in phase space. 
The regular trajectories show whispering-gallery motion and do not
cross the small quarter circle of radius $a$. Each trajectory which crosses this
curve is chaotic, see Fig.~\ref{fig:billiards:mushroom:general}(c). 
There is no hierarchical regular-to-chaotic transition 
region and there appear no resonance chains inside the regular island.
Hence, for this billiard resonance-assisted tunneling does not occur and 
the direct regular-to-chaotic tunneling process 
is relevant for all energies $E$.
The application of the fictitious integrable system approach
to the mushroom billiard leads to the explicit analytical formula
\eqref{eq:billiards:mushroom:gamma_final}, 
which was obtained in Ref.~\cite{BaeKetLoeRobVidHoeKuhSto2008},
where it was successfully compared to experimental data.

\subsubsection{Derivation of tunneling rates}

In order to predict tunneling rates for the mushroom billiard we 
review the derivation \cite{BaeKetLoeRobVidHoeKuhSto2008} starting from 
Eqs.~\eqref{eq:deriv:coupmatelorth_bil} and \eqref{eq:deriv:FGR_bil}.
First we construct a fictitious integrable system $\Hreg$,
determine its eigenstates $\psireg^{mn}(\bq)$, and find a model for the 
chaotic states $\psich(\bq)$. In the following analysis we set $R=1$.
A natural choice for the regular system $\Hreg$ is the quarter-circle billiard. 
Its eigenfunctions are analytically known 
\begin{eqnarray}
\label{eq:billiards:mushroom:psireg}
 \psireg^{mn}(r,\varphi) = N_{mn}J_m(j_{mn}r)\sin(m\varphi),
\end{eqnarray}
in polar coordinates $(r,\varphi)$. They are characterized by the radial 
$(n=1,2,\dots)$ and the azimuthal $(m=2,4,\dots)$ quantum numbers. As we are
considering the quarter-circle billiard $m$ is allowed to take even values only.
Here $J_m$ denotes the $m$th Bessel function, $j_{mn}$ the $n$th root of $J_m$,
$N_{mn}=\sqrt{8/\pi}/J_{m-1}(j_{mn})$ accounts for the normalization, and
$E_{mn}=j_{mn}^{2}$ are the eigenenergies.
Among the regular states of the quarter-circle billiard we will consider
only those which concentrate on regular tori of the
mushroom billiard with angular momentum $p_{mn}=m/j_{mn}>a$.
 
We use the Hamiltonian $H$ of the mushroom and $\Hreg$ of the 
quarter-circle billiard in Eq.~\eqref{eq:deriv:coupmatelorth_bil} 
to determine the coupling between the regular and the chaotic states. 
An infinite potential difference $H-\Hreg=-\infty$ occurs, while 
$\psireg^{mn}=0$ in the stem of the mushroom for $y<0$. 
In order to avoid the undefined product
$(H-\Hreg)\psireg^{mn}$ we introduce a finite potential at $y\leq 0$, see 
Fig.~\ref{fig:billiards:mushroom:general}(b),
\begin{eqnarray}
\label{eq:billiards:mushroom:HregW}
 \Hreg^W(\bq,\bp) & = & \bp^2 + V(\bq)\\
 V(\bq) & = & \left\{\begin{array}{ll} 0& \quad \text{for } x^2+y^2\leq 1,\; x,\,y>0\\ 
                                       W& \quad \text{for } y\leq 0,\; 0\leq x \leq 1\\
                                       \infty& \quad \text{otherwise}\end{array}\right. 
\end{eqnarray}
and consider the limit $W\to\infty$ in which the quarter-circle billiard is
recovered. For finite $W$ the regular eigenfunctions $\psiregW^{mn}(\bq)$ of
$\Hreg^W$ decay into the region $y<0$. 
To describe this decay we make the following ansatz
\begin{eqnarray}
 \psiregW^{mn}(x,y) = \psiregW^{mn}(x,y=0)\ue^{\lambda y},
\end{eqnarray}
where $\lambda$ depends on $W$ via the Schr\"odinger equation as
\begin{eqnarray}
\label{eq:billiards:mushroom:HregW_SGL}
 -\lambda^2 + W = E_{mn}^{W}.
\end{eqnarray}
Since the regular eigenfunctions $\psiregW^{mn}$ and their derivatives have to 
be continuous at $y=0$ we obtain
\begin{eqnarray}
 \lambda = \frac{\partial_y \psiregW^{mn}(x,y=0)}{\psiregW^{mn}(x,y=0)}.
\end{eqnarray}
Evaluating Eq.~\eqref{eq:deriv:coupmatelorth_bil} for the
coupling matrix elements one finds
\begin{eqnarray}
\label{eq:billiards:mushroom:v_final1}
 \vchmn & \hspace*{-0.05cm}= & \hspace*{-0.1cm}\lim_{W\to\infty} \Int_{0}^{a}
          \hspace*{-0.1cm}\ud x \Int_{-l}^{0}\hspace*{-0.1cm} \ud y\,\psich(x,y)
           (-W)\psiregW^{mn}(x,y)\quad\;\;\\
\label{eq:billiards:mushroom:v_final1b}
        & \hspace*{-2.0cm}= & \hspace*{-1.0cm}-\lim_{W\to\infty} 
                              \Int_{0}^{a}\ud x \Int_{-l}^{0} 
               \ud y\,\psich(x,y)\frac{W}{\lambda}\ue^{\lambda y}
               \partial_y\psiregW^{mn}(x,0)
\end{eqnarray}
In the last equation the term $W\ue^{\lambda y}/\lambda$ appears, which 
in the limit $W\to\infty$ gives for $y \leq 0$
\begin{eqnarray}
 \frac{W}{\lambda}\ue^{\lambda y} = \frac{W}{\sqrt{W-E_{mn}^{W}}}
                  \ue^{\sqrt{W-E_{mn}^{W}} y} \to 2\delta(y),
\end{eqnarray}
where we use Eq.~\eqref{eq:billiards:mushroom:HregW_SGL} and that $E_{mn}^{W}$ 
remains bounded. For the coupling matrix elements, 
Eq.~\eqref{eq:billiards:mushroom:v_final1b}, we obtain
\begin{eqnarray}
\label{eq:billiards:mushroom:v_final2}
 \vchmn & = & -\Int_{0}^{a}\ud x\,\psich(x,y=0)\partial_y\psireg^{mn}(x,y=0).\quad
\end{eqnarray} 
Due to the limiting process only an integration along the line $y=0$ remains
which connects the quarter circle billiard to the stem of the mushroom. 
Equation~\eqref{eq:billiards:mushroom:v_final2} contains the derivative
of the regular wave function perpendicular to this line. 
The largest contribution of the integral is close to the corner of
the mushroom at $x=a$, as the derivative of the regular eigenfunctions 
$\partial_y\psireg^{mn}$ decays toward $x=0$.
Inserting the regular states, Eq.~\eqref{eq:billiards:mushroom:psireg}, 
one finds
\begin{eqnarray}
\label{eq:billiards:mushroom:v_final3}
  \vchmn & = & -N_{mn}\Int_{0}^{a}\ud x\,\psich(x,y=0)\frac{m}{x}J_m(j_{mn}x).
             \quad
\end{eqnarray}

In order to evaluate Eq.~\eqref{eq:billiards:mushroom:v_final3} we 
use a random wave description to model the chaotic states $\psich(\bq)$. 
It has to respect the Dirichlet boundary conditions in the vicinity of 
the corner at $x=a$. For this random wave model we use polar coordinates 
$\bq=(\rho,\vartheta)$ as introduced in Fig.~\ref{fig:billiards:mushroom:general}(a) 
such that the corner of angle $3\pi/2$ is located
at $(0,0)$. The Dirichlet boundary conditions at this corner are accounted for
using \cite{Leh1959} 
\begin{eqnarray}
\label{eq:billiards:mushroom:psich}
 \psich(\rho,\vartheta) = \sqrt{\frac{8}{3\Achb}}\sum_{s=1}^{\infty} 
                          c_s J_{\frac{2s}{3}}(k\rho)
                          \sin\left(\frac{2s}{3}\vartheta\right)
\end{eqnarray}
in which $x-a=\rho\cos(\vartheta)$ and $y=\rho\sin(\vartheta)$. The coefficients
$c_s$ are independent Gaussian random variables with mean zero, 
$\langle c_s\rangle=0$, and unit variance, $\langle c_s c_t \rangle = \delta_{st}$.
Equation~\eqref{eq:billiards:mushroom:psich} fulfills the Schr\"odinger 
equation at energy $E=k^2$ and the prefactor is chosen such that 
$\langle|\psich(\rho,\vartheta)|^2\rangle = 1/\Achb$ holds far away from the 
corner. Note, that we do not require these chaotic states to decay into the 
regular island, as Eq.~\eqref{eq:billiards:mushroom:v_final3} is an integral
along a line of the billiard where the phase space is fully chaotic. 
Near the boundary, but far away from the corner, $\langle|\psich|^2\rangle$ 
recovers the behavior $1-J_0(2k|x|)$ \cite{BaeSchSti1998,Ber2002}.
Inserting Eq.~\eqref{eq:billiards:mushroom:psich} into
Eq.~\eqref{eq:billiards:mushroom:v_final3} at energy $E_{mn} = j_{mn}^{2}$ and 
$\vartheta=\pi$, we obtain the coupling matrix elements
\begin{eqnarray}
 \vchmn & = & -N_{mn}\sqrt{\frac{8}{3\Achb}} \sum_{s=1}^{\infty} c_s 
              \sin\left(\frac{2s}{3}\pi\right) \nonumber\\
      & & \cdot\Int_{0}^{a} J_{\frac{2s}{3}}(j_{mn}[a-x])\frac{m}{x} 
          J_{m}(j_{mn}x)\,\ud x,
\end{eqnarray}
where terms with $s$ a multiple of $3$ vanish. Using these matrix elements in 
Fermi's golden rule, Eq.~\eqref{eq:deriv:FGR_bil},
gives a prediction of the tunneling rates 
\begin{equation}
\label{eq:billiards:mushroom:gamma_final0}
 \gamma_{mn} = m^2N_{mn}^{2} \sideset{}{^\prime}\sum_{s=1}^{\infty}
  \left[ \Int_{0}^{a} \frac{\ud x}{x}\,
         J_{\frac{2s}{3}}(j_{mn}(a-x))J_m(j_{mn}x)\right]^2\hspace*{-0.2cm}.
\end{equation}
The prime at the summation indicates that the sum over $s$ excludes all
multiples of three. The remaining integral can be solved analytically 
\cite[Eq.~11.3.40]{AbrSte1970}, leading to the final result
\begin{eqnarray}
\label{eq:billiards:mushroom:gamma_final}
 \gamma_{mn} = \frac{8}{\pi} \sideset{}{^\prime}\sum_{s=1}^{\infty}
 \frac{J_{m+\frac{2s}{3}}(j_{mn} a)^2}{J_{m-1}(j_{mn})^2}.
\end{eqnarray}
This gives a prediction of direct regular-to-chaotic tunneling rates of
any regular state $\psireg^{mn}$ to the chaotic sea in the mushroom
billiard. The sum has its dominant contribution for $s=1$ and evaluating
Eq.~\eqref{eq:billiards:mushroom:gamma_final} up to 
$s\leq 2$ gives sufficiently accurate predictions.

It is worth to remark that a very plausible estimate of the
tunneling rate is given by the averaged square of the regular wave
function on a circle with radius $a$, i.e.\ the boundary to the
fully chaotic phase space, yielding $\gamma_{mn}^{0} = N_{mn}^{2}
J_{m}(j_{mn}a)^2/2$. 
Surprisingly, it is just about a factor of
$2$ larger for the parameters we studied. 
In Ref.~\cite{BarBet2007}
a related quantity is proposed, given by the integral of the squared
regular wave function over the quarter circle with radius $a$. This
quantity, however, is too small by a factor of order $100$ for the
parameters under consideration.

\subsubsection{Comparison with numerical rates}

\begin{figure}[tb]
  \begin{center}
    \includegraphics[width=8.5cm]{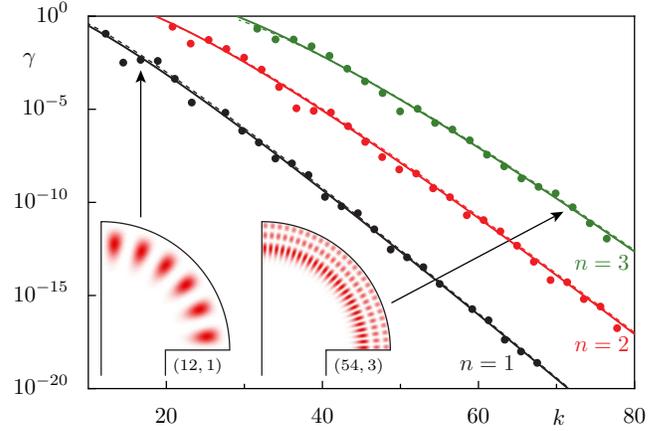}
    \caption{(Color online) Tunneling rates of regular states with
              quantum numbers $n \leq 3 $ vs $k$ for $a=0.5$ 
              comparing the prediction of Eq.~\eqref{eq:billiards:mushroom:gamma_final} 
              (solid lines) and numerical rates (dots). 
              The insets show the regular eigenfunctions $\psireg^{12,1}$ and
              $\psireg^{54,3}$ (as indicated by labels). 
              In addition the asymptotic prediction
              of Eq.~\eqref{eq:billiards:mushroom:approx4} is
              presented (dashed lines).
            }
    \label{fig:billiards:mushroom:res:var_m}
  \end{center}
\end{figure}

The eigenvalues and eigenfunctions of the mushroom billiard are determined by
numerically solving the Schr\"odinger equation.
The improved method of particular solutions
\cite{BetTre2005,BarBet2007} allows a determination of 
the energies $E$ with a relative error $\approx 10^{-14}$.
We analyze the widths $\Delta E_{\text{ch},mn}$ of avoided crossings 
between a given regular state and typically $30$ chaotic states under variation 
of the height $l$ of the stem, starting with $l=0.3$.
From Eq.~\eqref{eq:numerics:avcr}, 
$\gamma \approx \langle |\Delta E_{\text{ch},mn}|^2 \Achb\rangle / 8$, we deduce 
the tunneling rate where we use $\Achb=la+[\arcsin(a)+a\sqrt{1-a^2}]/2$ 
\cite{BaeKetLoeRobVidHoeKuhSto2008} as 
derived in Appendix~\ref{sec:appendix:ach}. 
Note, that some pairs of regular states are very close
in energy, e.g., $E_{20,1}-E_{16,2}\approx 10^{-4}$, such that their avoided
crossings with a chaotic state overlap, making a numerical
determination of the smaller tunneling rate unfeasible within
the presented approach. 

Figure~\ref{fig:billiards:mushroom:res:var_m} shows the numerical 
tunneling rates $\gamma_{mn}$
for fixed radial quantum number $n=1,2,3$ and increasing azimuthal quantum
number $m$ for $a=0.5$. It is compared to the theoretical prediction,
Eq.~\eqref{eq:billiards:mushroom:gamma_final}, which is connected for fixed $n$
and increasing $m$ by straight lines, giving rise to an apparently smooth curve.
We find excellent
agreement for tunneling rates $\gamma_{mn}$ over $18$ orders of magnitude.
The small oscillations which appear in the numerical rates on top of the
exponential decay might be related to the two-level approximation for avoided crossings 
which we use for the numerical determination of the tunneling rates. 

\begin{figure}[tb]
  \begin{center}
    \includegraphics[width=8.5cm]{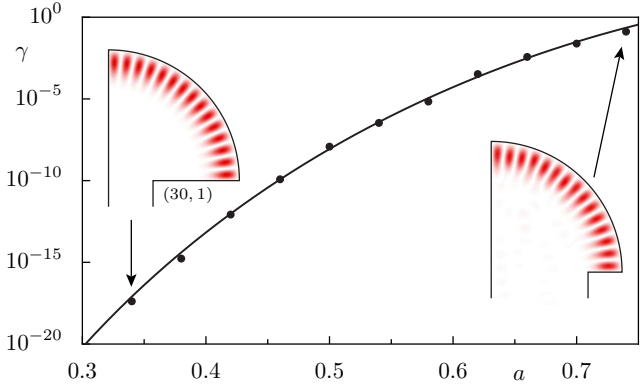}
    \caption{(Color online) Tunneling rates of the regular state $\psireg^{30,1}$ vs 
              the stem width $a$. We compare the prediction of 
              Eq.~\eqref{eq:billiards:mushroom:gamma_final} 
              (solid lines) and numerical rates (dots). 
              The insets show the regular eigenfunction $\psireg^{30,1}$ at
              $a=0.34$ and $a=0.74$.
            }
    \label{fig:billiards:mushroom:res:var_a}
  \end{center}
\end{figure}

To further test the prediction we determine the tunneling rate 
of the regular state $(m,n)=(30,1)$ under variation of the stem width $a$. 
The results presented in Fig.~\ref{fig:billiards:mushroom:res:var_a} 
show a decrease of this tunneling rate which 
appears faster than exponential with $1-a$. Again we find excellent 
agreement to numerical rates. Note, that the accuracy of the
numerical method used for determining eigenenergies of the mushroom is best 
for $a\approx 0.5$ and declines for larger or smaller $a$. 

\begin{figure}[tb]
  \begin{center}
    \includegraphics[width=8.5cm]{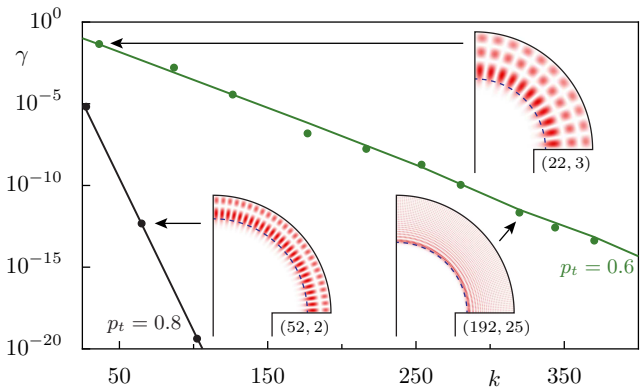}
    \caption{(Color online) Tunneling rates of regular states localized closest to
              a classical torus of angular momentum $p_t=0.8$ and $p_t=0.6$
              vs $k$ for $a=0.5$. We compare the prediction of 
              Eq.~\eqref{eq:billiards:mushroom:gamma_final} 
              (solid lines) and numerical rates (dots). 
              The insets show the regular eigenfunctions $\psireg^{52,2}$
              close to $p_t=0.8$ and $\psireg^{22,3}$,
              $\psireg^{192,25}$ close to $p_t=0.6$.
            }
    \label{fig:billiards:mushroom:res:var_torus}
  \end{center}
\end{figure}

Another interesting question is how the tunneling rates from a given classical 
torus behave. For this we consider a sequence of regular states $(m,n)$ which 
semiclassically localize on a torus characterized by an angular momentum 
$p_t > a$. For each $n$ we choose
$m$ such that $p_{mn} = m/j_{mn} \approx p_t$.
The resulting behavior of the tunneling rates is presented
in Fig.~\ref{fig:billiards:mushroom:res:var_torus} for $p_t=0.6$ and $p_t=0.8$. 
A comparison of these predictions to numerical 
rates shows excellent agreement.

\begin{figure}[tb]
  \begin{center}
    \includegraphics[width=8.5cm]{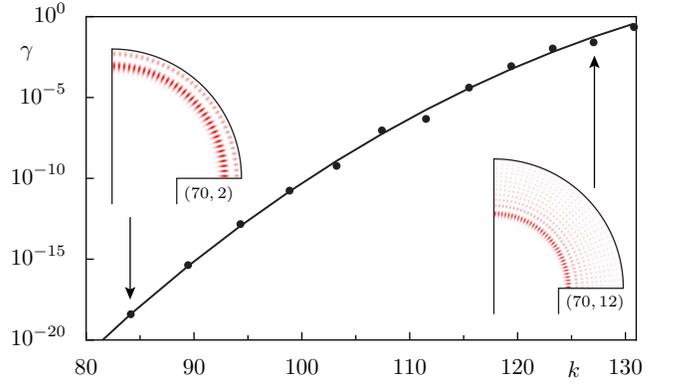}
    \caption{(Color online) Tunneling rates of regular states with
              quantum number $m = 70$ vs $k$ for $a=0.5$ 
              comparing the prediction of Eq.~\eqref{eq:billiards:mushroom:gamma_final} 
              (solid lines) and numerical rates (dots). 
              The insets show the regular eigenfunctions $\psireg^{70,2}$ and
              $\psireg^{70,12}$.
            }
    \label{fig:billiards:mushroom:res:var_n}
  \end{center}
\end{figure}

For fixed azimuthal quantum number $m$ and increasing radial quantum number $n$ 
the tunneling rates increase. This behavior is presented
in Fig.~\ref{fig:billiards:mushroom:res:var_n}. Again we find good agreement between
the predictions of Eq.~\eqref{eq:billiards:mushroom:gamma_final} and 
numerical rates. 

\subsubsection{Approximation}

Let us now approximate Eq.~\eqref{eq:billiards:mushroom:gamma_final} 
for $m \gg n$ and large wave numbers $k$ in order to understand 
the exponential behavior of the tunneling rates, which is visible in  
Figs.~\ref{fig:billiards:mushroom:res:var_m} and 
\ref{fig:billiards:mushroom:res:var_torus} with increasing $k$.
First we consider the numerator of the leading term $s=1$ 
in Eq.~\eqref{eq:billiards:mushroom:gamma_final} 
and use Ref.~\cite[Eq.~9.1.63]{AbrSte1970} 
for non-integer arguments of the Bessel function 
\begin{equation}
\label{eq:billiards:mushroom:approx1}
 J_{m+\frac{2}{3}}(j_{mn}a) \leq \left\vert a_{mn}^{m+\frac{2}{3}}
 \frac{\exp\left(\left(m+\frac{2}{3}\right)\sqrt{1-a_{mn}^{2}}\right)}
 {\left(1+\sqrt{1-a_{mn}^{2}}\right)^{m+\frac{2}{3}}}\right\vert.
\end{equation} 
with $a_{mn} = j_{mn}a/(m+2/3)$. Equation~\eqref{eq:billiards:mushroom:approx1} 
provides an upper bound of $J_{m+2/3}(j_{mn}a)$.
Numerically it has been confirmed, that a good approximation is given by
this bound divided by $m^{3/2}$,
\begin{eqnarray}
\label{eq:billiards:mushroom:approx2}
 J_{m+\frac{2}{3}}(j_{mn}a) \approx \frac{1}{m^{\frac{2}{3}}} \left( 
 a_{mn}^{\tilde{m}} \frac{\exp\left(\tilde{m}b_{mn}\right)}
 {\left(1+b_{mn}\right)^{\tilde{m}}} \right)&&\\
\label{eq:billiards:mushroom:approx2b}
    = \frac{1}{m^{\frac{2}{3}}}\exp\left(\tilde{m}\left[b_{mn} - 
          \ln\left(\frac{1+b_{mn}}{a_{mn}}\right)\right]\right), &&
\end{eqnarray} 
where $\tilde{m}=m+2/3$ and $b_{mn} = \sqrt{1-a_{mn}^{2}}$. 
The denominator of the term $s=1$ in Eq.~\eqref{eq:billiards:mushroom:gamma_final} 
can be approximated for $n=1$ using Ref.~\cite[Eq.~9.5.18]{AbrSte1970}
\begin{equation}
\label{eq:billiards:mushroom:approx3}
 J_{m-1}(j_{mn}) = J'_{m}(j_{mn}) \approx -1.1131\,m^{-\frac{2}{3}} 
 \approx -\frac{1}{m^{\frac{2}{3}}}.
\end{equation}
We thus obtain for the tunneling rates assuming that 
Eq.~\eqref{eq:billiards:mushroom:approx3} also 
approximately holds for $n>1$
\begin{equation}
\label{eq:billiards:mushroom:approx4}
 \gamma_{mn} \approx \frac{8}{\pi}\exp\left(2\tilde{m}\left[b_{mn} - 
          \ln\left(\frac{1+b_{mn}}{a_{mn}}\right)\right]\right).
\end{equation}
For fixed radial quantum number $n$ and increasing azimuthal 
quantum number $m$ the tunneling rates decay 
exponentially with $k \propto m$. Figure~\ref{fig:billiards:mushroom:res:var_m} shows the 
comparison to Eq.~\eqref{eq:billiards:mushroom:gamma_final} (dashed lines). 
We find agreement with deviations smaller than a factor of two.
The prediction, Eq.~\eqref{eq:billiards:mushroom:approx4}, has a similar form as
Eq.~\eqref{eq:maps:sc:wkb_formula} which has been 
obtained for a quantum map with
a harmonic oscillator-like regular island. This reflects the similarity of
this system to the mushroom billiard.

\subsection{Annular billiard}
\label{sec:billiards:annular}

\begin{figure}[tb]
  \begin{center}
    \includegraphics[width=8.5cm]{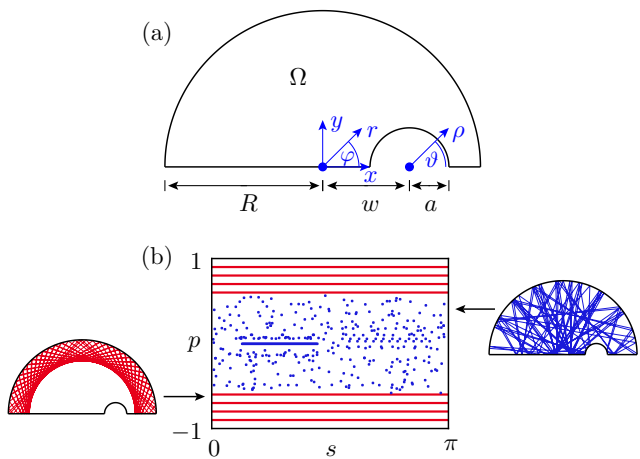}
    \caption{(Color online) (a) Schematic picture of the desymmetrized annular billiard
             showing the two coordinate systems used in the theoretical derivation
             of the direct regular-to-chaotic tunneling rates.
             (b) Poincar\'e section at the semi-circle boundary 
             (relative tangential momentum $p$ vs arclength $s$) 
             showing regular and chaotic regions for $R=1$, $a=0.15$, and
             $w=0.45$ with illustrations of trajectories.
            }
    \label{fig:billiards:annular:general}
  \end{center}
\end{figure}

We consider the desymmetrized annular billiard characterized
by the radius $R=1$ of the large semi-circle, the radius $a$ of the small 
semi-circle, and the displacement $w$ of this semicircle, 
see Fig.~\ref{fig:billiards:annular:general}(a). 
In contrast to the sharply separated regular and chaotic dynamics in the 
mushroom billiard, the phase space of the annular billiard is more subtle. 
Again we find regions of regular whispering-gallery motion. In the chaotic 
region, however, additional regular islands and partial barriers 
may be located, depending on the choice of $a$ and $w$. 
This structure leads to the existence of so-called beach states which 
resemble regular states but are 
localized in the chaotic sea close to the border of the regular region. 
These beach states were described by Doron and Frischat 
\cite{DorFri1995,FriDor1998} who also studied the dynamical tunneling process in 
annular billiards. Their prediction of 
tunneling rates required fitting with a free parameter. 
In this section we want to apply the fictitious integrable system approach 
to find a prediction of direct regular-to-chaotic tunneling rates which describe 
the decay of whispering-gallery modes into the chaotic sea.

In order to determine these direct regular-to-chaotic tunneling rates
we proceed similar to the case of the mushroom billiard. 
We have to evaluate Eq.~\eqref{eq:deriv:coupmatelorth_bil} for
the coupling matrix elements $\vchmn$ and then use Fermi's golden rule,
Eq.~\eqref{eq:deriv:FGR_bil}, to determine the tunneling rates.
In the first step the fictitious integrable system $\Hreg$ and its eigenstates 
$\psireg^{mn}$ have to be defined. A natural 
choice for $\Hreg$ is the semi-circle billiard which 
exactly reproduces the whispering-gallery motion in the annular billiard.
Its eigenstates are given by
\begin{eqnarray}
\label{eq:billiards:annular:psireg}
 \psireg^{mn}(r,\varphi) = N_{mn} J_m\left(j_{mn} r\right)\sin(m\varphi)
\end{eqnarray}
in polar coordinates $(r,\varphi)$, where $J_m$ denotes the $m$th Bessel 
function, $j_{mn}$ is the $n$th root of  $J_m$, and 
$N_{mn}=\sqrt{4/\pi}/J_{m-1}(j_{mn})$. 
The regular states are characterized by the radial quantum number $n=1,2,\dots$ 
and the azimuthal quantum number $m=1,2,\dots\;$. Hence, the tunneling rates 
describing the decay of the regular state $\psireg^{mn}(r,\varphi)$ will 
be labeled by $\gamma_{mn}$. Note, that for the annular billiard only those 
regular states semiclassically exist which localize on tori with
angular momentum $p_{mn} = m/j_{mn}>w+a$.

Evaluating Eq.~\eqref{eq:deriv:coupmatelorth_bil} in order to 
determine the coupling matrix elements $\vchmn$ between the regular and the 
chaotic states, an infinite potential difference arises within the small disk
of radius $a$ between the Hamiltonian $H$ of the annular and 
$\Hreg$ of the semi-circle billiard, $H-\Hreg=\infty$. 
At the same time for the chaotic states $\psich=0$ holds in 
that region, which leads to an
undefined product ``$\infty\cdot 0$''. Similar to the approach presented for 
the mushroom billiard we circumvent this problem by 
considering a finite potential difference $W$ for which at the end the limit
$W\to\infty$ is performed. 
In contrast to the mushroom billiard the area of the annular billiard $\Omega$
is included in the area of the semi-circle billiard 
$\Omegareg$, $\Omega\subset\Omegareg$. 
We find that in this case the derivative of the chaotic states 
$\psich(\bq)$ enters in Eq.~\eqref{eq:deriv:coupmatelorth_bil}. We obtain
\begin{eqnarray}
\label{eq:billiards:annular:v_final1}
  \vchmn  & = & a\Int_{0}^{\pi} \ud \vartheta\,
           \psireg^{mn}(a,\vartheta)\, \partial_{\rho}\psich(\rho=a,\vartheta)\\
\label{eq:billiards:annular:v_final2}
     & = & a N_{mn} \Int_{0}^{\pi} \ud \vartheta \, 
           J_m(j_{mn}r) \sin(m\varphi) \partial_{\rho}\psich(a,\vartheta) ,\quad\;\;\;
\end{eqnarray}
where we introduce polar coordinates $(\rho,\vartheta)$ such that 
$x = \rho\cos(\vartheta)+w$ and $y=\rho\sin(\vartheta)$.
Only the integration over $\vartheta$ from $\vartheta=0$ to $\vartheta=\pi$ along 
the small semi-circle of radius $\rho=a$ remains. 
Note, that the regular eigenfunctions are 
given in polar coordinates $(r(\rho,\vartheta),\varphi(\rho,\vartheta))$, 
while we integrate along $\rho=a$, see Fig.~\ref{fig:billiards:annular:general}(a).
Along this half-circle the regular wave function is 
largest near the point $(\rho,\vartheta) = (a,0)$ for $w>0$.
Hence, a random wave description for the chaotic
states $\psich(\rho,\vartheta)$ has to respect the Dirichlet boundary 
conditions on the line $y=0$ and on the small semi-circle of radius $a$.

Such a random wave model can be constructed using the solutions of the annular 
concentric billiard \cite[\S 25]{Som1978} as base functions in polar 
coordinates $(\rho,\vartheta)$
\begin{eqnarray}
\label{eq:billiards:annular:RWM}
 \psich(\rho,\vartheta) & = & \frac{2}{\sqrt{\Achb}} \sum_{s=1}^{\infty} 
                              c_s\sin(s\vartheta) \nonumber\\
    & & \cdot \frac{J_s(k\rho)Y_{s}(ka)-
              J_{s}(ka)Y_s(k\rho)}{\sqrt{J_{s}(ka)^2+Y_{s}(ka)^2}},
\end{eqnarray}
in which the $c_s$ are Gaussian random variables with 
$\langle c_s\rangle = 0$ and $\langle c_s c_t\rangle = \delta_{st}$. Furthermore, 
$Y_s$ denotes the $s$th Bessel function of the second kind. 
The normalization constant $2$ has been obtained numerically, such that 
$\langle|\psich|^2\rangle=1/\Achb$ holds far away from the boundary. 
The chaotic states defined by Eq.~\eqref{eq:billiards:annular:RWM} 
fulfill the Schr\"odinger equation at arbitrary energy $E=k^2$. 
Similar to the mushroom billiard we do not require that the chaotic states 
decay into the regular island, as 
Eq.~\eqref{eq:billiards:annular:v_final2} is an integral along a line 
of the billiard which is not hit by any regular whispering-gallery trajectory.
Near the horizontal boundary and away from the small circle 
$\langle|\psich|^2\rangle$ recovers the behavior 
$1-J_0(2 k |\rho-a|)$ \cite{BaeSchSti1998,Ber2002}.
Using the radial derivative of $\psich(\rho,\vartheta)$ at $\rho=a$ one finds
\begin{equation}
 \partial_{\rho}\psich(\rho=a,\vartheta) = \frac{2k}{\sqrt{\Achb}} 
  \sum_{s=1}^{\infty} c_s\sin(s\vartheta) R_s(ka),
\end{equation}
in which we use $J_{s+1}(ka)Y_{s}(ka)-J_{s}(ka)Y_{s+1}(ka) = 2/(\pi ka)$ 
\cite[Eq.~9.1.16]{AbrSte1970} and introduce
\begin{equation}
 R_s(ka) := \frac{2}{\pi ka \sqrt{J_{s}(ka)^2+Y_{s}(ka)^2}}.
\end{equation}
With this result we calculate the tunneling rates at energy $E_{mn}=j_{mn}^2$
\begin{eqnarray}
 \label{eq:billiards:annular:gamma_final}
 \gamma_{mn} & = & 2 N_{mn}^{2} a^2 j_{mn}^{2} \sum_{s=1}^{\infty}
      R_s(j_{mn}a)^2  \nonumber\\
      & & \cdot\left[ \Int_{0}^{\pi} \ud \vartheta \sin(s\vartheta)
          J_m(j_{mn}r(\vartheta))\sin(m\varphi(\vartheta))\right]^2
          \hspace*{-0.2cm}.\quad\;
\end{eqnarray}
This is our final result predicting the decay of a regular state $\psireg^{mn}$
located in the regular whispering-gallery region to the chaotic sea.
In Eq.~\eqref{eq:billiards:annular:gamma_final}, in contrast to the 
result for the mushroom billiard, Eq.~\eqref{eq:billiards:mushroom:gamma_final}, 
the term $s=1$ is typically not the most 
important contribution. Here the dominant $s$ increases with energy.
In contrast to the prediction of Refs.~\cite{DorFri1995,FriDor1998} 
no fitting is required.

The numerical determination of tunneling rates using avoided crossings is more
difficult for the annular billiard than in the case of the mushroom billiard:
In order to affect the chaotic
but not the regular component of phase space under parameter variation 
we increase the radius $a$ of the small inner circle and move its center 
position $w$ such that its rightmost edge at $w+a$ is constant.
We increase $a$ starting from $a=0.05$ until $30$ avoided crossings have occurred
or $a=0.4$ is reached.
This procedure drastically affects the chaotic states while the regular 
whispering-gallery modes remain almost unchanged. 
Note, that under this parameter variation the phase-space structure in the chaotic
region changes from macroscopically chaotic to a situation in which
additional regular islands and partial barriers appear.
With increasing $k$, however, a smaller parameter variation is required and 
these problems become less relevant.

\begin{figure}[tb]
  \begin{center}
    \includegraphics[width=8.5cm]{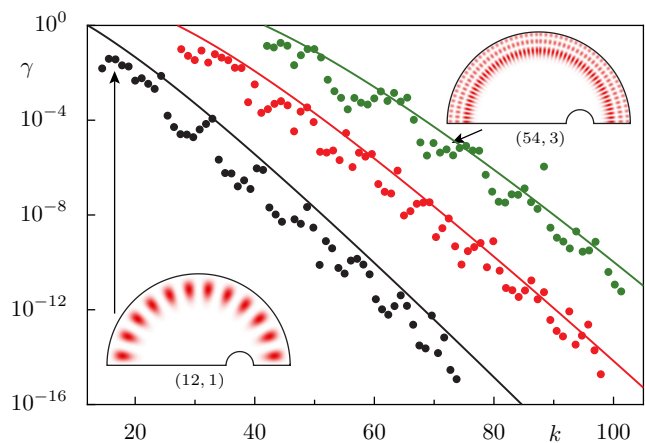}
    \caption{(Color online) Tunneling rates from regular states with
              quantum numbers $n \leq 3$ vs $k$ for $w+a=0.6$ 
              comparing the prediction of 
              Eq.~\eqref{eq:billiards:annular:gamma_final} 
              (solid lines) and numerical rates (dots). 
              The insets show the regular eigenfunctions $\psireg^{12,1}$ and
              $\psireg^{54,3}$.}
    \label{fig:billiards:annular:res:var_m}
  \end{center}
\end{figure}

Figure~\ref{fig:billiards:annular:res:var_m} shows the numerical tunneling 
rates for $w+a=0.6$ and the analytical prediction, 
Eq.~\eqref{eq:billiards:annular:gamma_final}, for $a=0.15$ and $w=0.45$.
Variations of $a$ with constant $w+a$, as necessary for the numerical rates,
affect the prediction by at most a factor of two (not shown).
The comparison in Fig.~\ref{fig:billiards:annular:res:var_m} gives
qualitative agreement. While some tunneling rates agree with the prediction,
deviations of a factor of $100$ appear for other rates.

We believe that these deviations are artifacts of our numerical procedure to 
determine tunneling rates from avoided crossings \cite{Loe2010}.
They occur most likely due to beach states which exist in the chaotic region 
of phase space and look similar to regular states though no quantizing tori 
can be associated with them \cite{DorFri1995,FriDor1998}. 
Numerically we need to analyze avoided crossings 
between the regular mode and all modes outside the regular island under 
variation of the shape of the billiard. However, as
beach states almost behave like regular states, their energy only slightly 
varies if the boundary of the billiard is changed. Hence, they rarely show 
avoided crossings with the regular state. In addition the chaotic states 
concentrate further away from the regular island and thus give rise to 
considerably smaller avoided crossings. This leads to artificially reduced 
numerical tunneling rates.
It would be desirable to determine numerical 
tunneling rates for the annular billiard
by other means, e.g., by opening the billiard, to obtain a more quantitative
verification of the prediction Eq.~\eqref{eq:billiards:annular:gamma_final}.

\subsection{Disordered wire in a magnetic field}
\label{sec:billiards:wire}

\begin{figure}[tb]
  \begin{center}
    \includegraphics[width=8.5cm]{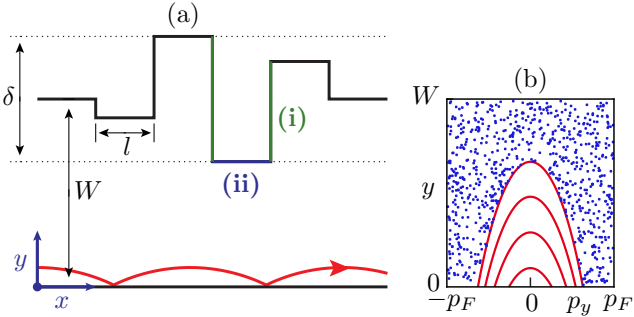}
    \caption{(Color online) (a) Schematic picture of a 
             magnetic wire with one-sided disorder
             to which leads of width $W$ are attached. The wire is composed of
             rectangular elements of length $l$ and a uniformly distributed 
             height in $[W-\delta/2, W+\delta/2]$.
             For the tunneling process contributions (i) along the vertical 
             and (ii) along the horizontal parts of the upper boundary
             are relevant. (b) A Poincar\'e section $(y,p_y)$ 
             at fixed $x$ shows a large regular island (red lines) 
             and the chaotic sea (blue dots).
            }
    \label{fig:billiards:wire:general}
  \end{center}
\end{figure}

We study two-dimensional nanowires 
\cite{GarGovWoe2002,FeiBaeKetRotHucBur2006,FeiBaeKetBurRot2009} 
with one-sided disorder, see Fig.~\ref{fig:billiards:wire:general}(a), 
in the presence of a homogeneous magnetic field $B$ perpendicular to the wire. 
Such nanowires have a mixed phase space, 
see Fig.~\ref{fig:billiards:wire:general}(b). 
Orbits which only hit the lower flat boundary are regular skipping orbits while 
those which are reflected at the upper disordered boundary numerically show 
chaotic motion. The phase space of the wire has a sharp transition from regular 
to chaotic dynamics. There are no resonance chains inside the regular 
island and there is no relevant hierarchical region. 

In Refs.~\cite{FeiBaeKetRotHucBur2006,FeiBaeKetBurRot2009} it was shown 
that in such wires the localization lengths $\xi$ increase exponentially
with increasing Fermi wave number $k_F$ of the electrons. This can be explained 
by the mixed phase-space structure giving rise to dynamical tunneling between 
the regular island and the chaotic sea. The dynamical tunneling rates $\gamma$ 
are directly related to the localization lengths, $\xi\propto 1/\gamma$.
We apply the fictitious integrable system approach to determine
localization lengths $\xi$ and compare the results to the analytical
prediction derived in Refs.~\cite{FeiBaeKetRotHucBur2006,FeiBaeKetBurRot2009}.

The nanowire with a disordered boundary is assembled from $L$ rectangular 
elements to which two leads of width $W$ are attached. The $n$th element has 
a length $l=W/5$ and a height $h_n$ which is uniformly distributed in the 
interval $[W-\delta/2, W+\delta/2]$ with $\delta=2W/3$. The heights $h_n$ 
are allowed to take $M=20$ different values including the boundaries of 
the interval. For increasing values of the Fermi 
wave number $k_F$ the magnetic field $B$ is adjusted such that the cyclotron 
radius $r_c = \hbar k_F/(eB)$ remains constant, using $r_c=3W$. 
This leaves the classical dynamics unchanged and the semiclassical 
limit is given by $k_F\to\infty$.

In order to apply the fictitious integrable system approach 
we have to consider a closed billiard. It is obtained by imposing periodic 
boundary conditions at the positions where the 
leads are attached. The tunneling rate is given by Fermi's golden rule, 
Eq.~\eqref{eq:deriv:FGR_bil}, using the coupling-matrix 
elements between regular and chaotic states, 
Eq.~\eqref{eq:deriv:coupmatelorth_bil}.
Similar to the examples discussed previously we have to find a fictitious 
integrable system $\Hreg$ which resembles the regular part of the phase space 
and extends it beyond. A reasonable choice is a billiard of length $Ll$ with 
periodic boundary conditions at $x=0$ and with 
a boundary at $y=0$ but open for $y>0$.
In a magnetic field it shows exactly the same regular phase space as the wire 
with one disordered boundary and is completely integrable.
Its eigenfunctions $\psireg^m(x,y)$ are given by \cite{Bal1999}
\begin{eqnarray}
 \psireg^m(x,y)=\frac{1}{\sqrt{lL}} \ue^{\ui \km x} Z_m(y),
\end{eqnarray}
where $Z_m(y)$ solves a one-dimensional Schr\"odinger equation with an 
effective potential due to the magnetic field and $m \geq 1$ is the quantum 
number in transversal direction.
Using this regular 
system and its eigenstates in Eq.~\eqref{eq:deriv:coupmatelorth_bil} gives 
\begin{eqnarray}
\label{eq:billiards:wire:v_0_part}
 \vchm = \Int_{\mathcal{C}} \ud s\, \partial_N\psich(x(s),y(s))\,\psireg^m(x(s),y(s)).
\end{eqnarray}
As for the annular billiard the derivative of the chaotic states 
$\partial_N \psich(x(s),y(s))$ is in the normal direction along the
disordered boundary $\mathcal{C}$, which is parameterized by $s$. 

The most important contributions of the integral in 
Eq.~\eqref{eq:billiards:wire:v_0_part} are from rectangular elements of the 
wire which have the smallest height, i.e.\ $h_n=W-\delta/2$. This follows from 
the exponential decay of the regular wave functions $\psireg^m$ in 
$y$-direction. There are $L/M$ such rectangular elements. The upper boundary
of the wire near such an element is composed of (i) two vertical parts
($x=nl$, $y\in[h_{n-1},W-\delta/2]$ and $x=(n+1)l$, $y\in[W-\delta/2,h_{n+1}]$)
and (ii) one horizontal part ($x\in[nl,(n+1)l]$, $y=W-\delta/2$).

(i) Let us first consider one vertical contribution to the coupling-matrix elements $\vchm$
for the fixed $x$-coordinate $x^*=nl$. 
For the chaotic states $\psich$ we employ a random wave model 
with wave length $k_F$ such that the Dirichlet boundary condition at $x=x^*$ is 
fulfilled
\begin{eqnarray}
\label{eq:billiards:wire:v_1_part_rwm}
 \psich(r,\varphi)=\frac{2}{\sqrt{\Achb}} \sum_{s=1}^{\infty} 
                   c_s J_s(k_F r)\sin\left(s\varphi-\frac{\pi}{2}s\right),
\end{eqnarray}
where the polar coordinates $(r,\varphi)$ are defined by $x=x^*+r\cos\varphi$ 
and $y=W-\delta/2+r\sin\varphi$.
The coefficients $c_s$ are Gaussian random variables with mean zero and 
$\langle c_s c_t \rangle = \delta_{st}$. 
It will turn out that the form of Eq.~\eqref{eq:billiards:wire:v_1_part_rwm}
is more convenient for the following evaluation than a plane wave ansatz.
The derivative of $\psich$ with respect to $x$ at $x=x^*$ gives
\begin{eqnarray}
 \partial_x \psich(x=x^*,\yh)=\frac{2}{\sqrt{\Achb}} 
                         \sum_{s=1}^{\infty} c_s \frac{J_s(k_F \yh)s}{\yh},
\end{eqnarray}
where $\yh=y-(W-\delta/2)$.
Hence, the contribution of one vertical part of the boundary to the coupling-matrix 
element is (up to a phase $\ue^{\ui\km x^*}$)
\begin{eqnarray}
\label{eq:billiards:wire:v_1_part}
 \vchm = \frac{2}{\sqrt{\Achb}}\frac{1}{\sqrt{Ll}} \sum_{s=1}^{\infty} c_s
  \Int_{0}^{\infty} \frac{J_s(k_F \yh) s}{\yh} Z_m(\yh) \, \ud \yh. \quad
\end{eqnarray}
Here we have replaced the upper integration limit by $\infty$ as $Z_m(\yh)$
decays exponentially. The sum of $2L/M$ such coupling matrix elements gives 
according to Fermi's golden rule, Eq.~\eqref{eq:deriv:FGR_bil}, the direct
regular-to-chaotic tunneling contribution of the vertical boundaries
\begin{eqnarray}
 \label{eq:billiards:wire:rate_part_1}
 \gamma_{m}^{\text{(i)}} = \frac{4}{Ml}\sum_{s=1}^{\infty} 
                          s^2 \left( \Int_{0}^{\infty} \frac{J_s(k_F \yh)}{\yh} 
          Z_m(\yh)\, \ud \yh\right)^2,
\end{eqnarray}
where we assume independent coefficients $c_s$ for each vertical boundary.

(ii) For the horizontal boundaries we consider the regular wave functions 
$\psireg^m(\xh,\yh=0)=\Zmn \ue^{\ui\km(nl+\xh)}$, where $\xh=x-nl$ and 
$\Zmn=Z_m(\yh=0)$.
For the chaotic wave function a random wave model respecting the Dirichlet 
boundary at $\yh=0$ is used, 
\begin{equation}
 \psich(\xh,\yh)= \sqrt{\frac{2}{N\Achb}} \sum_{s=1}^{N} c_s 
       \ue^{\ui(k_F\xh\cos\vartheta_s+\varphi_s)}\sin(k_F \yh \sin\vartheta_s),
\end{equation}
where the coefficients $c_s$ are Gaussian random variables with mean zero as well as 
$\langle c_s c_t \rangle = \delta_{st}$ and the angles $\vartheta_s$ and
$\varphi_s$ are uniformly distributed in $[0,2\pi)$.
The derivative of $\psich$ with respect to $\yh$ at $\yh=0$ reads
\begin{equation}
 \partial_{\yh}\psich(\xh,\yh=0)= \frac{\sqrt{2}k_F}{\sqrt{N\Achb}} \sum_{s=1}^{N} c_s 
       \sin\vartheta_s\, \ue^{\ui(k_F\xh\cos\vartheta_s+\varphi_s)}.
\end{equation}
Hence, we obtain for the contribution of one horizontal part of the boundary 
to the coupling-matrix element (up to a phase $\ue^{\ui\km nl}$)
\begin{equation}
\label{eq:billiards:wire:v_2_parta}
 \vchm = \frac{\sqrt{2}k_F \Zmn}{\sqrt{\Achb NLl}}
    \sum_s c_s \sin\vartheta_s\,\ue^{\ui\varphi_s}\hspace*{-0.07cm}\Int_{0}^{l}\ud \xh\,
    \ue^{\ui (k_F\cos\vartheta_s+\km) \xh}.
\end{equation}
The sum of $L/M$ such coupling matrix elements gives 
according to Fermi's golden rule, Eq.~\eqref{eq:deriv:FGR_bil}, the direct
regular-to-chaotic tunneling contribution of the horizontal boundaries
\begin{equation}
 \label{eq:billiards:wire:rate_part_2}
 \gamma_{m}^{\text{(ii)}} = \frac{|\Zmn|^2}{2\pi Ml}
    \Int_{0}^{2\pi} \ud\vartheta \sin^2\vartheta\, 
    \frac{2-2\cos\left(lk_F\left[\cos\vartheta+\frac{\km}{k_F}\right]\right)}
    {\left(\cos\vartheta+\frac{\km}{k_F}\right)^2},
\end{equation}
where we assume independent coefficients $c_s$ for each horizontal boundary.
The total tunneling rate is given as the sum of the two contributions, 
Eq.~\eqref{eq:billiards:wire:rate_part_1} from the vertical parts of the boundary and 
Eq.~\eqref{eq:billiards:wire:rate_part_2} from the horizontal parts of the boundary,
\begin{equation}
\label{eq:billiards:wire:rate_final}
 \gamma_m=\gamma_{m}^{\text{(i)}}+\gamma_{m}^{\text{(ii)}}.
\end{equation}
Here we use the approximation that random wave models for the horizontal and 
vertical boundaries are independent. 

\begin{figure}[tb]
  \begin{center}
    \includegraphics[width=8.5cm]{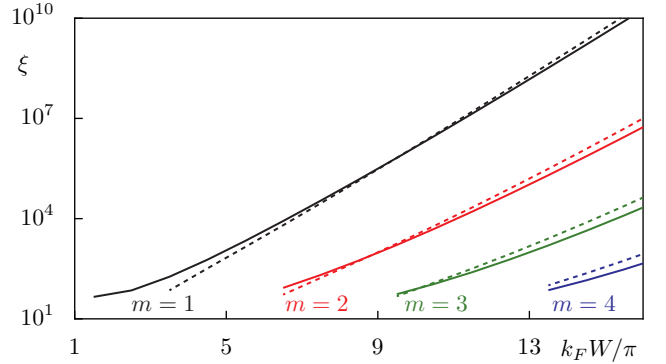}
    \caption{(Color online) Localization lengths of transversal modes in a magnetic wire with
             one-sided disorder for the mode numbers $m \leq 4$ vs $k_F W/\pi$ for
             $l/W=1/5$ and $\delta=2W/3$. 
             We compare the prediction of Eqs.~\eqref{eq:billiards:wire:rate_final} 
             and \eqref{eq:billiards:wire:rate_sc_wkb}
             (solid lines) and the analytical result of 
             Ref.~\cite{FeiBaeKetBurRot2009} (dashed lines).
            }
    \label{fig:billiards:wire:res:loclen}
  \end{center}
\end{figure}

We now compare this result for the dynamical tunneling rates $\gamma_m$
to the semiclassical prediction of localization lengths $\xi_m$ for the wire
with one-sided disorder derived in Ref.~\cite{FeiBaeKetBurRot2009}. 
For the connection between tunneling rates and localization lengths  we use
their inverse proportionality \cite{HufKetOttSch2002,IomFisZas2002}. One finds
\begin{eqnarray}
\label{eq:billiards:wire:rate_sc_wkb}
 \gamma_m = \frac{v_{x,m}}{\xi_m l} = \frac{2\km}{\xi_m l},
\end{eqnarray}
where $v_{x,m}$ is the velocity of the $m$th mode in $x$-direction
and the localization length $\xi_m$ is measured in units of the length $l$
of the rectangular elements.
Figure~\ref{fig:billiards:wire:res:loclen} 
compares the localization lengths obtained by the fictitious integrable system 
approach using Eqs.~\eqref{eq:billiards:wire:rate_final} and 
\eqref{eq:billiards:wire:rate_sc_wkb} with numerically determined $Z_m(y)$
(solid lines) to the analytical prediction of Ref.~\cite{FeiBaeKetBurRot2009}
(dashed lines) showing good agreement with deviations smaller than a factor of two.  

Note, that while in Ref.~\cite{FeiBaeKetBurRot2009} the localization length is 
$l$-independent we find for our horizontal contribution an $l$-dependence
due to the integral in Eq.~\eqref{eq:billiards:wire:v_2_parta}.
This becomes relevant for $l\gg W$ and can be corrected by either changing
the upper limit of integration in Eq.~\eqref{eq:billiards:wire:v_2_parta}
from $l$ to $1/k_F$ or by choosing an improved $\Hreg$. This accounts
for the fact that far away from the corners of the disordered boundary there 
is no tunneling.

\subsection{Optical microcavities}
\label{sec:cavities}

Optical microcavities in which photons can be confined in three
spatial dimensions are a subject of intensive research as they are relevant for
applications such as ultralow-threshold lasers.
Especially whispering-gallery cavities such as microdisks, microspheres, 
and microtoroids have been investigated as they can trap photons for a 
long time near the boundary by total internal reflection at the optically 
thinner medium. The corresponding whispering-gallery
modes have a very high quality factor $Q$.
While the microdisk emits the photons isotropically, cavities with deformed
surfaces may additionally lead to directed emission
\cite{NoeSto1997,WieHen2006,WieHen2008,YanWanDieHenWieYuPflBelYamKan2009}.

The quality factors $Q$ of optical microcavities can be connected to 
dynamical tunneling rates $\gamma$. These rates can be determined using the
fictitious integrable system approach \cite{BaeKetLoeWieHen2009},
which will be summarized in this section.
For a mode in an open cavity the quality factor $Q$ is related to
the corresponding resonance with
complex wave number $k=\real(k)+\ui\, \imag(k)$ via
\begin{equation} 
\label{eq:cavities:annular:eq:Q}
  Q=-\frac{\real(k)}{2\imag(k)}.
\end{equation}
For cavities with a mixed phase space the quality factor $Q$ of a regular mode
has two contributions
\begin{equation} 
\label{eq:cavities:annular:1/Q}
  \frac{1}{Q}  = \frac{1}{\Qod} + \frac{1}{\Qdyn}.
\end{equation}
Here $\Qod$ accounts for the coupling of the regular
mode to the continuum. In the case of the circular microcavity \cite{Noe1997} it is 
described by a quasi one-dimensional barrier-tunneling 
process through an angular momentum barrier.
It can be predicted by means of WKB theory or using the numerically 
determined decay rates of the modes in the circular cavity.
Note, that for this contribution the mixed phase-space structure is irrelevant.
The second contribution, $\Qdyn$, is given by dynamical tunneling
from the regular mode to the chaotic sea, which is
strongly coupled to the continuum.
Here we assume that there are no further phase-space structures within
the chaotic sea that affect the quality factor.
A priori it is not obvious, which of the two contributions will dominate.

In order to determine $\Qdyn$ the fictitious 
integrable system approach can be employed in analogy to hard-wall billiards.  
It is extended to open cavities in the
following way: (i) As a fictitious integrable system $\Hreg$ one chooses
a cavity such that it resembles
the regular dynamics of $H$. The quantum system has resonance states $\psireg$.
(ii) As a model for the
chaotic resonances $\psich$ a random wave model is used,
which in addition fulfills the relevant cavity boundary conditions.
(iii) The tunneling rate $\gamma$ determines the quality factor $\Qdyn$ by
\begin{equation} 
\label{eq:cavities:annular:Qdyn}
  \Qdyn  =  \frac{2 \real(k)^2 \no^2}{\gamma},
\end{equation}
where $\no$ is the refractive index of the cavity.
Here it is used that the imaginary part of the wave number is 
connected to the decay rate of the resonant state via 
$\gamma = -4 \no^{2} \real(k) \imag(k)$ and 
Eq.~\eqref{eq:cavities:annular:eq:Q}.

This approach was developed and applied to the annular 
microcavity, a microdisk with an air hole, in Ref.~\cite{BaeKetLoeWieHen2009}. 
It has a particularly interesting 
geometry which allows for unidirectional emission and high quality factors
simultaneously.
An analytical expression for the quality factor
$\Qdyn$ was derived which is in very good agreement with the
full numerical simulations of Maxwell's equations.


\section{Summary and outlook}
\label{sec:summary}

We study the direct regular-to-chaotic tunneling process in systems with 
a mixed phase space. It is dominant in the regime, 
$\heff \lesssim \Areg$, where fine-scale structures of the phase 
space such as nonlinear resonances are not resolved by quantum mechanics. 
To describe this tunneling process we introduce the fictitious integrable 
system approach. It uses a decomposition of the Hilbert space into two parts which
account for the regular and the chaotic dynamics. This leads to a formula
which predicts the direct tunneling rate $\gamma_m$ of the $m$th regular state
$|\psireg^m\rangle$ to the chaotic sea, Eq.~\eqref{eq:deriv:final_result}. 
The fictitious integrable system has to be chosen such that its dynamics
resembles the regular dynamics of the original mixed system as closely as 
possible and extends it beyond its regular region. The determination of 
the fictitious integrable system is the most difficult step in the 
application of this approach. For maps the Lie transformation or methods 
based on the frequency map analysis can be used. A general procedure 
for billiards is under development. 

In Sec.~\ref{sec:maps:app} dynamical tunneling rates are determined 
numerically and compared to the prediction of the fictitious integrable system 
approach for different quantum maps. We find excellent 
agreement over several orders of magnitude in $\gamma$. For a harmonic 
oscillator-like island embedded in a chaotic sea it is possible 
to derive a semiclassical expression for the tunneling rates which depends on 
the area of the regular region and the effective Planck constant only. 
Furthermore, in Sec.~\ref{sec:billiards} we apply 
the approach to billiard systems, where we use random wave 
models to describe the chaotic states. For the mushroom 
billiard the prediction is in excellent agreement with numerical 
and experimental data. 
Finally, we apply the approach for direct regular-to-chaotic
tunneling rates to the annular billiard, nanowires with one-sided disorder 
in a magnetic field, and to the quality factors of optical microcavities.

In the semiclassical regime, $\heff\ll\Areg$, nonlinear resonances lead to
enhanced tunneling rates due to resonance-assisted tunneling.
This resonance-assisted tunneling mechanism is combined to direct 
regular-to-chaotic tunneling in Ref.~\cite{LoeBaeKetSch2010}, leading to a 
prediction of tunneling rates which is valid from the quantum to the
semiclassical regime. For this prediction the direct tunneling rates 
discussed in this paper, which can be determined using the fictitious integrable
system approach, are essential.

Ultimately, it is the aim to obtain a complete semiclassical treatment of 
the approach also for systems with a generic regular island. 
This theory should predict tunneling rates
and their dependence on the effective Planck constant $\heff$
directly from classical properties of the regular island, such as its size,
shape, and winding number and properties of the chaotic sea, such as
unstable fixed points, partial barriers, and transport properties.
A promising approach seems to be the complex-path formalism
which has been successfully applied to study the tunneling tails of a
time-evolved wave packet in mixed regular-chaotic systems
\cite{ShuIke1995,ShuIke1998,OniShuIkeTak2001}.
Alternatively, a complex time approach \cite{Mou2007,DenMou2010}
can be considered. Using such semiclassical approaches for the prediction 
of regular-to-chaotic tunneling rates should give further insight 
into the dynamical tunneling process.

\begin{acknowledgments}
We are grateful to J.~Burgd\"orfer, S.~Creagh, J.~Feist, B.~Huckestein, 
S.~Fishman, M.~Hentschel, R.~H\"ohmann, A.~K\"ohler, U.~Kuhl, N.~Mertig, 
A.~Mouchet, M.~Robnik, S.~Rotter, P.~Schlagheck, A.~Shudo, H.-J.~St\"ockmann,
S.~Tomsovic, G.~Vidmar, and J.~Wiersig for valuable cooperations and 
stimulating discussions.
Furthermore, we acknowledge financial support through the DFG Forschergruppe 760 
``Scattering systems with complex dynamics''.
\end{acknowledgments}


\begin{appendix}

\section{Dimensionless Fermi's golden rule for time-periodic quantum systems}
\label{sec:appendix:fgr}

In order to predict dynamical tunneling rates $\gamma_m$ we use Fermi's golden 
rule for systems with a discrete spectrum. 
This rate describes the decay $\ue^{-\gamma_m t}$ 
of the $m$th regular state to the chaotic sea at most up to the Heisenberg time 
$\tau_H=\heff/\Delta_{\text{ch}}$.
Compared to the standard derivation of Fermi's golden rule \cite{Coh1995} for 
the continuous case, the incoherent integral over the continuum states is 
replaced by the incoherent sum over the discrete chaotic states. One finds
\begin{eqnarray}
\label{eq:app:fgr:1}
 \Gamma_m = \frac{2\pi}{\hbar} \langle|\Vchm|^2\rangle \rho_{\text{ch}}
\end{eqnarray}
with the chaotic density of states $\rho_{\text{ch}}$ and the average 
of the modulus squared of the coupling matrix 
elements $\Vchm$ of the Hamiltonian 
between energetically close-by chaotic states and the $m$th purely regular state.

We now apply Eq.~\eqref{eq:app:fgr:1} to time-periodic systems
with period $\tau$ and quasi-energies in the interval $[0,\hbar\omega]$
with $\omega=2\pi/\tau$ such that $\rho_{\text{ch}}=\Nch/(\hbar\omega)$.
We introduce dimensionless quantities $\gamma_m=\Gamma_m\tau$ and 
$\vchm = \Vchm\tau/\hbar$ leading to
\begin{eqnarray}
\label{eq:app:fgr:2}
 \gamma_m = N_{\text{ch} }\langle|\vchm|^2\rangle = 
           \sum_{\text{ch}} |\vchm|^2.
\end{eqnarray}
Here $\vchm=\langle\psich|U|\psireg^m\rangle$ is the coupling matrix element
of the time evolution operator $U$ over one period 
and the mean level spacing is $2\pi/N$.

\section{Adapted eigenstates of the harmonic oscillator}
\label{sec:appendix:ho_states}

For the maps $\mapd$, which for $R=0$ show a
tilted and squeezed harmonic oscillator-like regular island,
purely regular states $\psireg^m(q)$ can be constructed by
tilting and squeezing the eigenfunctions of the harmonic oscillator accordingly. 
One finds \cite{Sch2006}
\begin{equation}
\label{eq:app:quant:ef_ho}
 \psireg^m(q) = \frac{1}{\sqrt{2^m m!}} 
                \left(\frac{\real(\sigma)}{\pi\hbareff}\right)^{\frac{1}{4}}
    H_m\left(\sqrt{\frac{\real(\sigma)}{\hbareff}}q\right) 
                \ue^{-\frac{\sigma}{\hbareff}\frac{q^2}{2}}
\end{equation}
with the Hermite polynomials $H_m$ and the complex squeezing parameter 
$\sigma$, which can be obtained from the tilting angle $\theta$ of the 
elliptic island with respect to the momentum axis and the ratio $\omega$ 
of its half axes with
\begin{eqnarray}
 \real(\sigma) & = & \frac{1}{\frac{1}{\omega}\cos^2(\theta)+\omega\sin^2(\theta)}, \\
 \imag(\sigma) & = & \real(\sigma)\left(\frac{1}{\omega}-\omega\right)\sin(\theta)\cos(\theta).
\end{eqnarray}
Here $\theta$ and $\omega$ can be determined from the linearized 
dynamics of the classical map around the fixed point,
\begin{eqnarray}
\label{eq:app:maps:lin_theta}
 \theta & = & \frac{1}{2}\arctan\left(\frac{\monmat_{22}-
               \monmat_{11}}{\monmat_{12}-\monmat_{21}}\right),\\
 \omega & = & \sqrt{\frac{|\monmat_{12}-\monmat_{21}|-c}
               {|\monmat_{12}-\monmat_{21}|+c}},\quad
\end{eqnarray}
where $c = \sqrt{(\monmat_{12}+\monmat_{21})^2+(\monmat_{22}-\monmat_{11})^2}$
and 
\begin{equation}
 \monmat = \left(\begin{array}{cc}
            \tfrac{\partial q_{n+1}}{\partial q_{n}} & 
              \tfrac{\partial q_{n+1}}{\partial p_{n}} \\
	    \tfrac{\partial p_{n+1}}{\partial q_{n}} & 
               \tfrac{\partial p_{n+1}}{\partial p_{n}}
           \end{array}\right).
\end{equation}

\section{Derivation of $\Achb$ for the mushroom billiard}
\label{sec:appendix:ach}

\begin{figure}[tb]
\begin{center}
\includegraphics[]{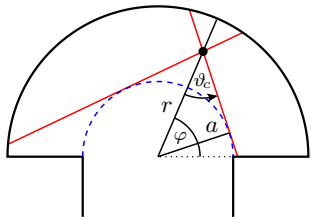}
\caption{(Color online) All trajectories of the mushroom billiard passing 
          through the point $(r,\varphi)$ with $r>a$ (dot) and angle 
          $|\vartheta|<\vartheta_c$ cross the half-circle of radius $a$ 
          (dashed line) and are chaotic. For angles 
          $|\vartheta+\pi|<\vartheta_c$ the time-reversed trajectories cross the 
          half-circle and they are chaotic as well.}
          \label{fig:app:mushroom_ach}
\end{center}
\end{figure}

We want to show how the area $\Achb$, which is the area of the billiard times the 
fraction of the chaotic phase-space volume, is determined 
for the mushroom billiard.

All trajectories which enter the stem of the mushroom or cross the half-circle
$r=a$ in the cap are chaotic irrespective of their momentum direction. 
Additionally the region $r>a$ contains 
some chaotic dynamics. We denote its contribution to $\Achb$
by $I$ and obtain for the mushroom billiard
\begin{equation}
\label{eq:app:Ach:1}
 \Achb = 2la + \frac{\pi}{2}a^2 + I.
\end{equation}
In order to determine $I$ we consider the fraction
$P(r)$ of chaotic trajectories in the cap for $r>a$. It 
depends on the radial coordinate $r$ only
\begin{equation}
\label{eq:app:Ach:2}
 P(r) = \frac{2}{\pi}\arcsin\left(\frac{a}{r}\right),
\end{equation}
as can be seen from Fig.~\ref{fig:app:mushroom_ach}.

Integrating over the region $r>a$ in the cap of the mushroom gives
\begin{eqnarray}
 I & = & \Int_{0}^{\pi} \Int_{a}^{R} r \,\ud r \,\ud\varphi\, P(r)\\
   & = & R^2 \arcsin\left(\frac{a}{R}\right) + a \sqrt{R^2-a^2}-\frac{1}{2} 
         \pi a^2.
\end{eqnarray}
Inserting this expression into Eq.~\eqref{eq:app:Ach:1} we finally obtain
\begin{equation}
\label{eq:app:Ach:result}
 \Achb = 2la + \left[R^2\arcsin\left(\frac{a}{R}\right)+a\sqrt{R^2-a^2}
                 \right]
\end{equation}
as the the area of the mushroom billiard times the 
fraction of the chaotic phase-space volume. For the desymmetrized mushroom 
used in Sec.~\ref{sec:billiards:mushroom} this result has to be
divided by two.

\end{appendix}


\end{document}